\documentclass[
 reprint,
 superscriptaddress,
 footinbib,
 nobibnotes,
 amsmath,amssymb,
 aps,
 prx,
]{revtex4-2}

\usepackage{booktabs}
\usepackage{graphicx}%
\usepackage{dcolumn}%
\usepackage{bm}%
\usepackage{siunitx}
\usepackage{xspace}
\usepackage{comment}
\usepackage{xcolor}
\DeclareSIUnit{\gauss}{G}

\newcommand{\di}{\mathrm d}
\newcommand{\ket}[1]{\ensuremath{\lvert #1 \rangle}\xspace}%
\newcommand{\bra}[1]{\ensuremath{\langle #1 \rvert}\xspace}%
\usepackage[colorlinks=True, citecolor=blue, urlcolor=blue, linkcolor=blue]{hyperref}

\makeatletter

\let\originaldoauthor\doauthor

\def\doauthor#1#2#3{%
  \begingroup
    \let\authornotes\@empty
    \def\equalcontribution{%
      \g@addto@macro\authornotes{%
        \comma@space
        \frontmatter@footnote{These authors contributed equally to this work.}%
      }%
    }%
    \originaldoauthor{#1}{#2}{#3\authornotes}%
  \endgroup
}

\makeatother

\renewcommand{\selectlanguage}[1]{}

\begin{document}

\newcommand{\partitle}[1]{\subsection*{#1}}

\title{Quantum phase estimation on a Rydberg quantum simulator}

\author{Hendrik Timme\equalcontribution}%
\email{hendrik.timme@mpq.mpg.de}
\affiliation{Max-Planck-Institut f\"{u}r Quantenoptik, 85748 Garching, Germany}
\affiliation{Munich Center for Quantum Science and Technology (MCQST), 80799 Munich, Germany}
\affiliation{Fakult\"{a}t f\"{u}r Physik, Ludwig-Maximilians-Universit\"{a}t, 80799 Munich, Germany}

\author{Benjamin F.~Schiffer\equalcontribution}%
\affiliation{Max-Planck-Institut f\"{u}r Quantenoptik, 85748 Garching, Germany}
\affiliation{Munich Center for Quantum Science and Technology (MCQST), 80799 Munich, Germany}
\affiliation{Collegium Helveticum, 8092 Zurich, Switzerland}
\affiliation{Institute for Theoretical Physics, ETH Zurich, 8093 Zurich, Switzerland}

\author{Ohad Lib}%
 \affiliation{Max-Planck-Institut f\"{u}r Quantenoptik, 85748 Garching, Germany}
  \affiliation{Munich Center for Quantum Science and Technology (MCQST), 80799 Munich, Germany}

\author{Shijia Sun}%
  \affiliation{Max-Planck-Institut f\"{u}r Quantenoptik, 85748 Garching, Germany}
  \affiliation{Munich Center for Quantum Science and Technology (MCQST), 80799 Munich, Germany}
  \affiliation{Fakult\"{a}t f\"{u}r Physik, Ludwig-Maximilians-Universit\"{a}t, 80799 Munich, Germany}

\author{Maximilian Ammenwerth}%
    \altaffiliation{Present address: JILA, University of Colorado Boulder, CO 80309, USA}
    \affiliation{Max-Planck-Institut f\"{u}r Quantenoptik, 85748 Garching, Germany}
    \affiliation{Munich Center for Quantum Science and Technology (MCQST), 80799 Munich, Germany}

\author{Dominik S.~Wild}%
    \altaffiliation{Present address: Google Quantum AI, 80636 Munich, Germany}
    \affiliation{Max-Planck-Institut f\"{u}r Quantenoptik, 85748 Garching, Germany}
    \affiliation{Munich Center for Quantum Science and Technology (MCQST), 80799 Munich, Germany}

\author{J.~Ignacio Cirac}%
  \affiliation{Max-Planck-Institut f\"{u}r Quantenoptik, 85748 Garching, Germany}
  \affiliation{Munich Center for Quantum Science and Technology (MCQST), 80799 Munich, Germany}

\author{Immanuel Bloch}%
  \affiliation{Max-Planck-Institut f\"{u}r Quantenoptik, 85748 Garching, Germany}
  \affiliation{Munich Center for Quantum Science and Technology (MCQST), 80799 Munich, Germany}
  \affiliation{Fakult\"{a}t f\"{u}r Physik, Ludwig-Maximilians-Universit\"{a}t, 80799 Munich, Germany}

\author{Johannes Zeiher}
    \email{johannes.zeiher@mpq.mpg.de}
    \affiliation{Fakult\"{a}t f\"{u}r Physik, Ludwig-Maximilians-Universit\"{a}t, 80799 Munich, Germany}
    \affiliation{Max-Planck-Institut f\"{u}r Quantenoptik, 85748 Garching, Germany}
    \affiliation{Munich Center for Quantum Science and Technology (MCQST), 80799 Munich, Germany}

\date{\today}

\begin{abstract}

Determining the energy spectrum of a many-body quantum system is a central goal of quantum physics.
Ground-state energies underpin materials science and quantum chemistry, whereas excited states govern dynamical phenomena such as quantum thermalization.
Quantum simulators~\cite{Cirac2012Goals,Gross2017Quantum,Altman2021Quantum} have greatly shaped our understanding of many-body systems, for example, by enabling access to microscopic mechanisms and nonlocal order parameters~\cite{Sherson2010Singleatomresolved, Bakr2010Probing, Schauss2015, Bernien2017Probing, Ebadi2021Quantum, Scholl2021Quantum}.
However, experiments still struggle to access the full range of many-body spectral features and are typically limited to measuring energy gaps and excitations by spectroscopic means~\cite{Senko2014Coherent,Landig2015Measuring,Manovitz2025Quantum,Chen2025Spectroscopy,Sun2026Observation}.
Here, we address this challenge by demonstrating quantum phase estimation on an analog Rydberg quantum simulator~\cite{Browaeys2020}.
We reconstruct the complex-valued return amplitude, or Loschmidt echo, of an initial probe state.
From its Fourier transform, we measure absolute ground-state energies and energy gaps in systems of up to 19 atoms in one and two spatial dimensions.
Varying the initial probe state, we select distinct spectral windows within which we compute the longitudinal and transverse magnetization, and observe their competition across the many-body spectrum.
Our results thereby demonstrate phase-sensitive many-body spectroscopy on an analog quantum simulator without auxiliary qubits or controlled evolution.
\end{abstract}

\maketitle

\begin{figure*}[t!]
    \centering
    \includegraphics[width=1\textwidth]{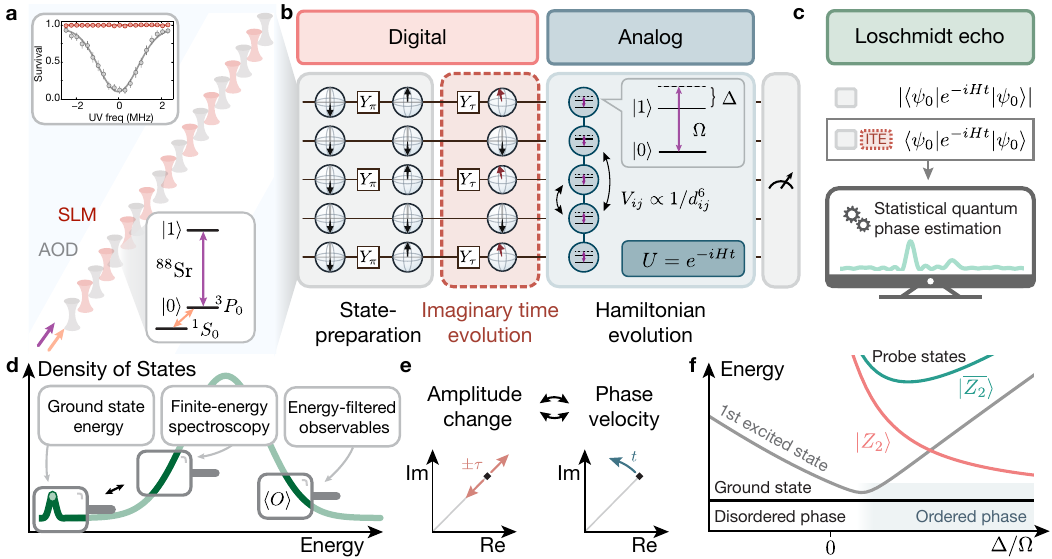}
    \caption{\textbf{Digital-analog quantum phase estimation on a Rydberg quantum simulator.}
    \textbf{(a)} Annotated fluorescence image of individual $^{88}\mathrm{Sr}$ atoms trapped in \SI{813}{\nano\meter} optical tweezers generated by a spatial light modulator (SLM) and crossed acousto-optic deflectors (AODs).
    Insets: level scheme of the clock--Rydberg qubit, and clock-to-Rydberg spectroscopy of atoms held in SLM traps (red) and of released AOD atoms (gray), showing the tweezer-induced light shift that provides local control.
    \textbf{(b)} Schematic of the hardware-efficient quantum phase estimation protocol.
    A main branch uses $\ket{Z_2}$ as the probe state, prepared via local control, while two imaginary-time evolution (ITE) branches include $Y_{\tau}$ rotations approximating the ITE step.
    All branches undergo the same analog evolution at positive detuning $\Delta$, followed by a computational-basis measurement.
    Counting outcomes equal to $\ket{Z_2}$ gives the return probability.
    \textbf{(c)} The three branches together determine the complex Loschmidt echo $\langle\psi_0|e^{-iHt}|\psi_0\rangle$ used in statistical QPE.
    \textbf{(d)} Fourier processing gives the probe-state-dependent density of states (DOS); extending the protocol to generalized Loschmidt echoes yields energy-filtered observables.
    \textbf{(e)} The Cauchy--Riemann equations relate the imaginary-time derivative of the magnitude to the real-time phase velocity of the Loschmidt echo.
    \textbf{(f)} Schematic of the ordered and disordered regimes of the one-dimensional chain and the spectral regions selected by two classical probes.
    }
    \label{fig:fig1}
\end{figure*}

Quantum phase estimation (QPE) is the main algorithmic method to probe spectral features on quantum computers~\cite{Abrams1999Quantum,Kitaev2002Classical}.
In standard implementations, QPE requires a controlled many-body time evolution dependent on the state of an auxiliary register~\cite{Cleve1998Quantum,Nielsen2010Quantum}.
However, such controlled dynamics are not native to analog simulators and their implementation on digital quantum computers requires deep compiled circuits.
To extract many-body spectral features even without this control, hardware-efficient approaches have been devised.
In one approach, the availability of special reference states with a known eigenenergy~\cite{Lu2021Filtering,Maskara2025Programmable} has been leveraged to experimentally extract many-body energies on trapped-ion and superconducting quantum processors~\cite{Hmery2024, Yoshioka2025Krylov}.
More generally, hardware-efficient approaches avoid the controlled evolution at the cost of additional measurements~\cite{Yang2024,Schiffer2025,Tabares2025Estimating,Clinton2026Quantum,Patel2026Quantum}.
The key concept in most of these approaches is the complex-valued \textit{Loschmidt echo}, which is the overlap of an initial state with its time-evolved copy.
Its Fourier transform yields the state-dependent density of states and enables statistical QPE and spectral filtering~\cite{Somma2019Quantum, OBrien2019Quantum, Ge2019Faster, Lu2021Filtering, Lin2022HeisenbergLimited,Dutkiewicz2022Heisenberglimited, Schuckert2023Probing,Blunt2023Statistical}.
Experimentally accessible return probabilities determine the magnitude of the Loschmidt echo, but lack information about its phase.
Recovering this missing phase without controlled many-body evolution is therefore a central experimental challenge.

Rydberg atom arrays have emerged as a particularly versatile platform for spectroscopic studies of many-body systems.
Trapped neutral atoms can be arranged in one- and two-dimensional geometries and coupled through strong, tunable Rydberg interactions, thereby realizing Ising-type spin models with site-resolved readout~\cite{Guardado-Sanchez2018Probing,Browaeys2020,Schauss2015,Bernien2017Probing,Ebadi2021Quantum,Scholl2021Quantum}.
In the blockaded regime, the constrained dynamics of these models have revealed quantum phase transitions~\cite{Keesling2019}, nonergodic revivals from $\mathbb{Z}_2$-ordered initial states~\cite{Turner2018,Bernien2017Probing}, and frustrated and topological states in two dimensions~\cite{Semeghini2021Probing}.
Recent experiments have accessed elementary excitations through quench dynamics in dipolar XY magnets~\cite{Chen2025Spectroscopy}.
Near criticality, coarsening dynamics have been linked to the low-energy spectrum~\cite{Manovitz2025Quantum}, while modulation spectroscopy has resolved finite-size conformal-field-theory spectra~\cite{Sun2026Observation}.
Single-hole injection into a doped antiferromagnet~\cite{Qiao2025} has also provided access to spatially resolved spectral functions of magnetic polarons~\cite{Martin2026}.
All of these results leverage the control achieved in analog quantum simulators to characterize transition energies and spectral weights, but do not provide access to the complex-valued many-body return amplitude.

In this work, we report the first experimental realization of statistical QPE on an analog quantum simulator, using the hardware-efficient protocol proposed in Refs.~\cite{Yang2024,Schiffer2025}.
We combine digital state preparation with native analog evolution to reconstruct the magnitude and phase of the Loschmidt echo.
Its Fourier transform gives the state-dependent density of states, from which we extract ground-state energies and the position of the first excited band in one-dimensional chains.
We access higher-energy spectral information by varying the digitally prepared probe states.
Beyond energy estimation, we extend the reconstruction to generalized Loschmidt echoes and obtain energy-filtered one- and two-point observables~\cite{Lu2021Filtering}.
We resolve the sign change of the transverse-drive contribution to the energy, which is a coherence-sensitive quantity that populations alone cannot provide.
Finally, we apply our method to a two-dimensional array and compute energy-resolved spatial density modes.
Together, these results demonstrate phase-sensitive measurements of interacting many-body systems on analog quantum simulators.
\begin{figure*}[t!]
    \centering
    \includegraphics[width=1\textwidth]{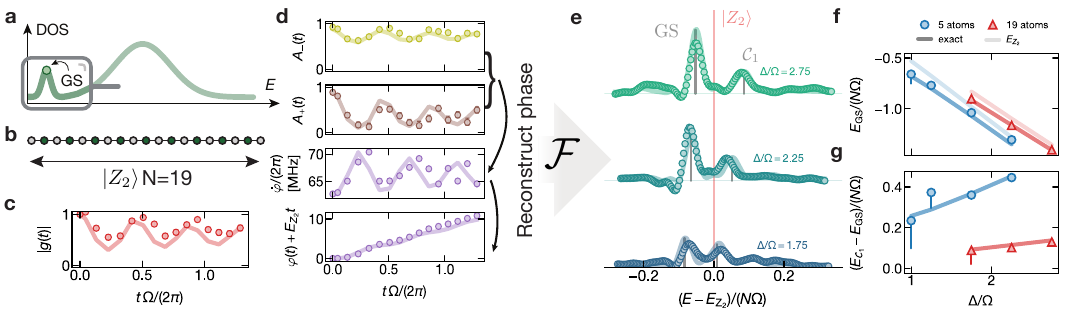}
    \caption{
    \textbf{Probing the ground-state energy via phase estimation.}
    \textbf{(a)}~Schematic of a Gaussian density of states (DOS) and a ground state (GS) peak.
    \textbf{(b)}~The $\ket{Z_2}$ N\'eel state in a chain of $N=19$ atoms; gray and green denote the Rydberg state $\ket{1}$ and clock state $\ket{0}$, respectively.
    \textbf{(c)}~Measured modulus of the Loschmidt echo after a quench as a function of the quench time.
    Markers show experimental data, normalized to the first point at $t=0$ to mitigate state preparation and readout errors, compared with numerical simulations (solid line).
    \textbf{(d)}~The upper panels show the normalized amplitudes of the ITE branches
    $A_-(t)$ and $A_+(t)$ used to reconstruct the phase velocity; see SI Sec.~\ref{app:postprocessing}.
    Below are the reconstructed phase velocity and integrated phase.
    The known mean energy $E_{Z_2}$ of the probe $\ket{Z_2}$ is removed in the displayed phase.
    The measured amplitudes correspond to the 19-atom chain at $\Delta/\Omega=2.25$.
    \textbf{(e)}~A Fourier transformation ($\mathcal{F}$) yields Gaussian-filtered spectra at three detunings, resolving the ground-state peak and the first excited band.
    The energy axis is shifted by the classical probe energy $E_{Z_2}$.
    \textbf{(f)}~Ground-state energy $E_{\rm GS}/(N\Omega)$ for five atoms
    (blue) and 19 atoms (red). Markers denote Gaussian fits to the experimentally reconstructed spectra, and vertical bars show central $68\%$ paired shot-bootstrap intervals.
    Solid curves show the exact numerical ground-state energies, while transparent curves show $E_{Z_2}/(N\Omega)$.
    \textbf{(g)}~Effective separation between the fitted ground-state and first-excited-band peaks. Vertical bars show central $68\%$ paired shot-bootstrap intervals for
    $(E_{\mathcal C_1}-E_{\rm GS})/(N\Omega)$.
    Solid curves show the separation between the exact ground state and the probe-weighted numerical mean energy of the first excited band.
    Details of the peak extraction and bootstrap procedure are given in SI Sec.~\ref{app:postprocessing}.}
    \label{fig:fig2}
\end{figure*}

\label{sec:qpe_rydbergs}
Our experiment features strontium-88 atoms at programmable positions $\mathbf r_i$ in one- and two-dimensional arrays~\cite{Norcia2018}.
The states $\ket{0}$ and $\ket{1}$ denote the clock (${}^3\mathrm{P}_0$) and Rydberg ($n=47$ ${}^3\mathrm{S}_1$) states, respectively, which are coupled by a laser with Rabi frequency $\Omega$ and detuning $\Delta$; see Fig.~\ref{fig:fig1}a.
The Hamiltonian describing this system can be written as
\begin{equation}
    H =
    \frac{\Omega}{2}\sum_{i=1}^{N}X_i
    -\Delta\sum_{i=1}^{N}n_i
    +\sum_{i<j}V_{ij}n_i n_j,
    \label{eq:rydberg_hamiltonian}
\end{equation}
where $X_i$ is a Pauli operator on site $i$, $n_i=\ket{1}_i\bra{1}_i$ projects onto the Rydberg state, and $V_{ij}=C_6/d_{ij}^6$ describes the van der Waals interaction, with coefficient $C_6$ and interatomic distance \mbox{$d_{ij}=|\mathbf r_i-\mathbf r_j|$}.
We take $\hbar=1$ and use angular-frequency units for $H$ and its eigenvalues throughout.
Setting the interatomic distance to \SI{2.6}{\micro\meter} corresponds to a nearest-neighbor (nn) interaction strength of $V_{\rm nn} / (2\pi) =\SI{35}{\mega\hertz}$.
At $\Omega/(2\pi)\approx 2.3$--$\SI{3.2}{\mega\hertz}$, we operate in the strongly blockaded regime, where the nearest-neighbor interaction satisfies $V_{\rm nn}\gg \Omega,|\Delta|$.
In this limit, the experiment approximately realizes a \textit{PXP model}, where simultaneous excitation of nearest-neighbor atoms is forbidden~\cite{Lesanovsky2012Interacting}, but finite longer-ranged interactions are retained.
The Hamiltonian is generically nonintegrable and features rich many-body physics that has been studied both in and out of equilibrium~\cite{Schauss2015,Bernien2017Probing, Keesling2019,Browaeys2020,Ebadi2021Quantum,Scholl2021Quantum,Semeghini2021Probing}.

A central quantity for statistical QPE is the Loschmidt echo.
The Loschmidt echo describes the complex-valued overlap of an initially prepared state $\ket{\psi_0}$ with its evolved state,
\begin{equation}
     g(t)=\bra{\psi_0}e^{-iHt}\ket{\psi_0}
    \equiv |g(t)|e^{i\varphi(t)}.
    \label{eq:loschmidt_amplitude}
\end{equation}
The modulus $|g(t)|$ is accessible in precision many-body quantum simulators~\cite{Jurcevic2017Direct,Xu2020Probing,Karch2025Probing} by extracting the overlap of a time-evolved state with an initially prepared state.
However, the modulus contains only information about energy differences~\cite{Kanno2025Tensorbased} and is generally insufficient to obtain the density of states.
Resolving eigenenergies of the Hamiltonian directly therefore also requires the phase $\varphi(t)$.

To reconstruct the full complex-valued $g(t)$ experimentally for the Rydberg Hamiltonian, Eq.~\eqref{eq:rydberg_hamiltonian}, we leverage a hardware-efficient hybrid protocol using digital and analog control~\cite{Schiffer2025}.
Specifically, we first use a layer of digital single-qubit rotations to prepare the N\'eel state $\ket{Z_2}=\ket{1010\ldots1}$ as an initial probe state for QPE; see Fig.~\ref{fig:fig1}b.
For a one-dimensional chain with an odd number of atoms, the N\'eel state represents a low-energy state in the ordered phase at $\Delta/\Omega>(\Delta/\Omega)_{\rm crit}$; see Fig.~\ref{fig:fig1}f.
After the digital preparation step, we subject the state to analog evolution $U=\exp(-iHt)$ for given Hamiltonian parameters $(\Omega, \Delta)$.
We then perform a computational-basis measurement by eliminating the Rydberg population through autoionization and imaging the remaining clock state population.
The fraction of bitstrings matching the initial probe yields the return probability $|g(t)|^2$; see Supplementary Information (SI) Secs.~\ref{app:experimental_platform} and~\ref{app:postprocessing} for details.

To recover the phase, we include a short imaginary-time evolution (ITE) step before the analog evolution~\cite{Yang2024,Schiffer2025}.
Although ITE is a nonunitary operation, we can precompile short normalized ITE for a known initial state into a unitary preparation pulse~\cite{Motta2020Determining}.
The ITE step $\pm\tau$ changes the magnitude $|g(t)|$ of the Loschmidt echo in the complex plane.
The Cauchy--Riemann equations then relate a change of $|g(t)|$ in imaginary-time evolution to a change of the phase of the Loschmidt echo, $\varphi(t)$, in real time; see Fig.~\ref{fig:fig1}e.
Integrating this phase velocity from the known initial value $\varphi(0)=0$ recovers the phase information of the full complex Loschmidt echo; see Fig.~\ref{fig:fig1}c and SI Sec.~\ref{app:ite_derivation}.
Crucially, for the N\'eel state in the PXP model, the nontrivial part of an approximate ITE rotates only single spins that are excited.
Small $Y$ rotations realize the normalized ITE of the $X$ drive to leading order.
The main and ITE branches then undergo the same native many-body evolution, without using auxiliary qubits or controlled many-body operations; see Fig.~\ref{fig:fig1}b and SI Sec.~\ref{app:ite_derivation}.

The Loschmidt echo provides access to spectral information of the quantum many-body model via the state-dependent density of states (DOS)~\footnote{The state-dependent density of states is also referred to as the local DOS in the literature. We avoid this term here to emphasize that \textit{local} does not refer to a position in real space.}; see Fig.~\ref{fig:fig1}d.
We define this spectrum as $\rho_{\psi_0}(E,\delta)\propto\bra{\psi_0}P_\delta(E)^2\ket{\psi_0}$, where we use a Gaussian spectral filter $P_\delta(E)=\exp[-(H-E)^2/(4\delta^2)]$, centered at energy $E$.
The DOS depends on the initial state and the filter width $\delta$, and is obtained by a Fourier transform of $g(t)$.
The Gaussian spectral filter is directly implemented as a Gaussian temporal filter on the time-evolved data; see SI Secs.~\ref{app:postprocessing} and~\ref{app:parameters}.
The maximum evolution time therefore limits the accessible energy resolution.
The mean energy of the initial state fixes the energy offset and is calculated from the Hamiltonian in Eq.~\eqref{eq:rydberg_hamiltonian}.

We first demonstrate the outlined protocol to extract the Loschmidt echo of a 19-atom chain initialized
in the $\ket{Z_2}$ N\'eel state, probing the ground-state energy and first excited band; see Fig.~\ref{fig:fig2}a,b.
The measured return-amplitude modulus exhibits periodic revivals, in good agreement with the model; see Fig.~\ref{fig:fig2}c.
The two ITE branches provide the phase needed to reconstruct the complex Loschmidt echo; see Fig.~\ref{fig:fig2}d.
At detunings $\Delta/\Omega\in\{1.75,2.25,2.75\}$, the resulting DOS exhibits a dominant low-energy peak; see Fig.~\ref{fig:fig2}e.
Its position, extracted from a Gaussian fit to the data, agrees with the numerical ground-state energy; see SI Sec.~\ref{app:postprocessing}.

For a quantitative analysis, we compare the ground-state energy $E_{\rm GS}$ with the mean probe-state energy $E_{Z_2}$ in Fig.~\ref{fig:fig2}f.
In the classical limit $\Delta/\Omega\to\infty$, the ground state of the projected model approaches the $\ket{Z_2}$ product state, so that $E_{\rm GS}\to E_{Z_2}$; in Fig.~\ref{fig:fig2}f the measured points approach the transparent $E_{Z_2}$ curves with increasing $\Delta/\Omega$.
Toward the critical region, quantum fluctuations generated by the projected transverse drive
$O_X=\sum_{i=1}^{N}P_{i-1}X_iP_{i+1}$, with $P_i=1-n_i$ and $P_0=P_{N+1}=1$, increasingly dress the product state and lower the ground-state energy below the energy of the $\ket{Z_2}$ state.
We can estimate this lowering with a perturbative approach, writing $H=H_0+H_X$ with $H_0=H_\Delta+H_V$ diagonal in the computational basis and $H_X=\Omega O_X/2$.
Each initially excited spin contributes one allowed virtual spin flip, so that neglecting the longer-ranged Rydberg interaction tails gives $E_{\rm GS}-E_{Z_2}\simeq-[(N+1)/2]\,\Omega^2/(4\Delta)$.
In addition, the initial state also has overlap with states belonging to the first excited band above the ground state.
In the classical limit, this band consists of configurations with one fewer Rydberg excitation than the N\'eel probe state.
Because the individual eigenstates in this band are not resolved experimentally, we fit the position of the corresponding peak in our data and compare its separation from the ground-state peak with the numerical separation $(E_{\mathcal C_1}-E_{\rm GS})/(N\Omega)$, using the mean band energy $E_{\mathcal C_1}$ weighted by the probe-state overlap; see SI Sec.~\ref{app:postprocessing}.
The results for 5 and 19 atoms agree well with the numerical predictions within the statistical uncertainties; see Fig.~\ref{fig:fig2}g.

\begin{figure}[t]
    \centering
    \includegraphics[width=1\columnwidth]{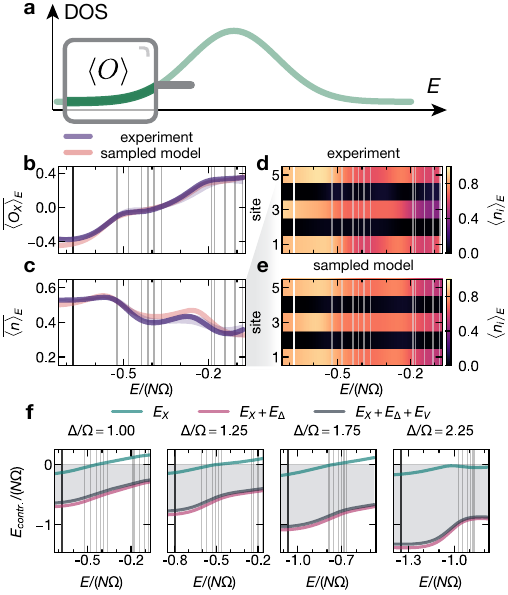}
    \caption{\textbf{Energy-filtered observables.}
    \textbf{(a)}~Schematic of a Gaussian filter selecting a finite energy window for evaluating an observable.
    \textbf{(b,c)}~Mean projected transverse coherence $\overline{\langle O_X\rangle}_E$ and mean Rydberg density $\overline{\langle n\rangle}_E$ in a five-atom chain at $\Delta/\Omega=1$.
    The transverse coherence changes sign from negative at low energies to positive at higher energies, distinguishing energy-lowering from energy-raising drive contributions.
    Dark purple curves show the experiment and pink curves the sampled model, evaluated on the experimental time grid and processed with the same finite-time filter.
    Shaded bands denote pointwise 68\% shot-bootstrap intervals.
    Vertical lines indicate exact ground-state and excited-state energies.
    \textbf{(d,e)}~Site-resolved filtered densities $\langle n_i\rangle_E$ from experiment and the sampled model, respectively, for the same parameters.
    \textbf{(f)}~Cumulative energy contributions $E_X$, $E_X+E_\Delta$, and $E_X+E_\Delta+E_V=\langle H\rangle_E$ at the respective detunings, where $E_X$, $E_\Delta$, and $E_V$ are drive, detuning, and interaction-tail contributions.
    Gray shading marks the total filtered energy relative to zero.
    }
    \label{fig:fig3}
\end{figure}

The lowering of the ground-state energy raises a microscopic question: how do different terms in the Hamiltonian compete in the spectrum?
To resolve these energy contributions, we compute energy-filtered observables~\cite{Lu2021Filtering}, building on the protocol used for statistical QPE.
We consider generalized Loschmidt echoes $g_b(t)=\bra{b}e^{-iHt}\ket{\psi_0}$, where $\ket{b}$ is a computational-basis state that can be represented by a bitstring.
Extracting these amplitudes therefore requires measuring the full counting statistics of those bitstring outcomes, which are accessible in precision Rydberg quantum simulators~\cite{Shaw2024Benchmarking}.
The integration constants of their phases, $\varphi_b(0^+)$, are fixed analytically in the PXP model; see SI Sec.~\ref{app:phase_anchor}.
Applying the spectral filter to each $g_b(t)$ then reconstructs the energy-filtered state $\ket{\psi_{E,\delta}}=\sum_b c_b(E)\ket{b}$, from which energy-resolved observables are evaluated; see Fig.~\ref{fig:fig3}a.
The filtering is performed in classical postprocessing so that the filtered state does not need to be prepared directly in the experiment.
From this state, we compute the energy-filtered observable $O_{\psi_0,\delta}(E) =\bra{\psi_{E,\delta}}O\ket{\psi_{E,\delta}}$, abbreviated as $\langle O\rangle_E$ below.

For the Rydberg array, we evaluate the Rydberg density $\overline{\langle n\rangle}_E=\sum_i\langle n_i\rangle_E/N$ and the transverse coherence $\overline{\langle O_X\rangle}_E=\sum_i\langle P_{i-1}X_iP_{i+1}\rangle_E/N$.
As the Rydberg density is diagonal in the measurement basis, it depends only on $|c_b(E)|^2$, whereas the transverse coherence also depends on the relative phases between the $c_b(E)$.
Our experimental data for both quantities are in good agreement with noiseless model dynamics evaluated on the experimental time grid and processed with the same finite-time filter; see Fig.~\ref{fig:fig3}b--e.
The observations can be explained intuitively:
For the Rydberg density, we find approximately discrete jumps from one spectral band to another.
This indicates that the different excitation bands predominantly differ by the number of Rydberg excitations.
For the transverse coherence, we find a similar stepwise behavior as a function of energy, but with a sign change from negative values to positive values.
In our convention we choose $\Omega>0$ so that negative $E_X(E)=(\Omega/2)\langle O_X\rangle_E$ corresponds to a drive contribution that effectively lowers the energy.
For the five-atom chain at $\Delta/\Omega=1$, the measured expectation value of the transverse drive is negative for the ground state, approximately zero in the first excited band and positive in the second excited band; see Fig.~\ref{fig:fig3}b.
Starting from the classical limit, the sign reversal can be understood in terms of level repulsion induced by the transverse drive.
Perturbatively, coupling to higher-energy configurations gives a negative drive contribution, whereas coupling to lower-energy configurations gives a positive contribution.
For an eigenstate with energy $E_n$, the Hellmann--Feynman theorem~\cite{Feynman1939} implies $\langle O_X\rangle_n=2\,\partial E_n/\partial\Omega$ at fixed detuning and interactions.
The ground-state contribution is negative, and since $O_X$ is traceless, positive contributions must occur in the excited spectrum.

The filtered observables also separate the contributions to the Hamiltonian $H = H_X + H_\Delta + H_V$.
We evaluate $E_X(E)=(\Omega/2)\langle O_X\rangle_E$, $E_\Delta(E)=-\Delta\sum_i\langle n_i\rangle_E$, and $E_V(E)=\sum_{i+1<j}V_{ij}\langle n_i n_j\rangle_E$, shown cumulatively in Fig.~\ref{fig:fig3}f.
Here, contributions from nearest-neighbor pairs vanish because those configurations are excluded by the blockade constraint.
We find that, across the four detunings in Fig.~\ref{fig:fig3}f, the diagonal contribution $E_\Delta + E_V$ dominates.
The transverse-drive contribution $E_X$ is smaller but not negligible, lowering the energy near the ground state and raising it in the excited bands.
Remarkably, the decomposition connects the measured spectral energies to their different density, interaction, and coherence contributions.

\begin{figure}[t]
    \centering
    \includegraphics[width=1\columnwidth]{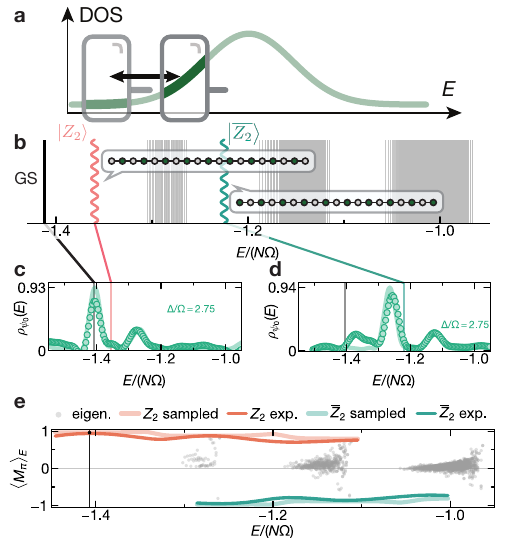}
    \caption{\label{fig:fig4} \textbf{Finite-energy many-body spectroscopy.}
    \textbf{(a)} Changing the initial probe selects different spectral weights.
    \textbf{(b)} Numerically computed eigenenergies of the one-dimensional Rydberg model at $\Delta/\Omega=2.75$ for $N=19$ atoms.
    The ground-state energy is shown in black and excited-state energies in gray.
    Wavy red and green lines indicate the mean energies of $\ket{Z_2}$ and $\ket{\overline{Z_2}}$, respectively.
    \textbf{(c)} The $\ket{Z_2}$-dependent spectrum resolves the ground-state sector and first excited band.
    \textbf{(d)} The higher-energy $\ket{\overline{Z_2}}$ probe emphasizes the first and second excited bands and shows probe-state-dependent spectral structure.
    \textbf{(e)} Filtered staggered magnetization $\langle M_\pi\rangle_E$ for the two probes, showing the retention of contrasting density-wave patterns.
    Gray markers show the numerically calculated magnetizations of the eigenstates. The black marker shows the magnetization of the ground state.
    Shaded uncertainty bands in (e) denote pointwise 68\% shot-bootstrap intervals.}
\end{figure}
So far, we have used the N\'eel state to probe the low-energy spectrum.
To access higher-energy bands, we change the digitally prepared initial state.
Furthermore, we examine how the microscopic structure of the filtered state depends on the probe state properties, using the same reconstruction protocol; see Fig.~\ref{fig:fig4}a.
For the 19-atom chains, we prepare the complementary N\'eel state $\ket{\overline{Z_2}}$, with excitations on even sites.
As it contains one fewer Rydberg excitation than $\ket{Z_2}$, this probe has a higher mean energy at the parameters considered here and shifts the sampled spectral weight toward higher energies; the computed spectrum and the mean energies of both probes are shown in Fig.~\ref{fig:fig4}b.

At $\Delta/\Omega=2.75$, the $\ket{Z_2}$ probe is dominated by the ground state and first excited band (Fig.~\ref{fig:fig4}c), whereas $\ket{\overline{Z_2}}$ carries appreciable weight across the first and second excited bands; see Fig.~\ref{fig:fig4}d.
In particular, the second band near $E/(N\Omega)\approx-1.12$ becomes visible in the reconstructed DOS, whereas it is largely absent for the $\ket{Z_2}$ probe.
We characterize the energy-filtered magnetization in Fig.~\ref{fig:fig4}e using the staggered magnetization
$M_\pi=N^{-1}\sum_{i=1}^{N}(-1)^{i-1}(2n_i-1)$.
The probe states $\ket{Z_2}$ and $\ket{\overline{Z_2}}$ have $M_\pi=+1$ and $-1$, respectively.
We find that the filtered states largely preserve the density-wave structure of their respective probes, while the average magnetization of the eigenstates clusters around zero.
A microcanonical ensemble would weight the eigenstates equally within an energy window; see also Refs.~\cite{DAlessio2016Quantum, Schuckert2025Observation}.

\begin{figure}[t]
    \centering
    \includegraphics[width=1\columnwidth]{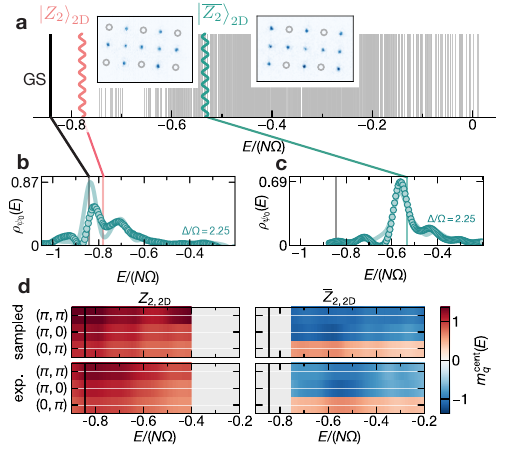}
    \caption{\label{fig:fig5} \textbf{Energy-resolved density order in 2D.}
    \textbf{(a)}~Numerically computed spectrum of the $5\times3$ Rydberg array and mean energies of the checkerboard-like probe state $\ket{Z_2}_{\rm 2D}$ and its counterpart $\ket{\overline{Z_2}}_{\rm 2D}$ at higher energy.
    Insets show fluorescence images of the two probe state configurations; empty circles mark sites whose atom was excited to the Rydberg state and therefore removed before imaging.
    \textbf{(b,c)}~Probe-state-dependent densities of states at $\Delta/\Omega=2.25$.
    The first probe emphasizes the ground-state region; the second shifts spectral weight to higher energies.
    \textbf{(d)}~Energy-resolved centered density modes $m_{\mathbf q}^{\mathrm{cent}}(E)$ at $\mathbf q=(\pi,\pi)$, $(\pi,0)$, and $(0,\pi)$ for the two probe states.
    The upper and lower rows show the sampled-model and experimental reconstructions, respectively.
    The $(\pi,\pi)$ mode probes checkerboard density order, while $(\pi,0)$ and $(0,\pi)$ probe stripe-like modulations.
    Positive and negative values distinguish opposite phases of each density modulation.
    Gray regions indicate energies outside the retained spectral support; the vertical black line marks the ground-state energy.}
\end{figure}

Finally, we extend our method to two dimensions and implement the protocols on a $5\times3$ square lattice at $\Delta/\Omega=2.25$.
Although its denser spectrum lacks the clearly separated excitation bands of the one-dimensional chains, see Fig.~\ref{fig:fig5}a, energy filtering still allows us to characterize the spatial structure of the probe-dependent filtered states.
We investigate this structure using density modes that distinguish checkerboard and stripe-like patterns.
We retain the nearest-neighbor blockade with finite longer-range interactions.
Because diagonal neighbors lie within the blockade radius, we additionally approximate the auxiliary preparation by rotating only initially excited sites.
We quantify the effect of this approximation on the reconstructed phase velocity in SI Sec.~\ref{app:2d_preparation_error}.
We use the two product-state probes $\ket{Z_2}_{\rm 2D}$ and $\ket{\overline{Z_2}}_{\rm 2D}$ shown in the insets in Fig.~\ref{fig:fig5}a.
Each outer row has the pattern $\ket{10101}$ in $\ket{Z_2}_{\rm 2D}$ and $\ket{01010}$ in $\ket{\overline{Z_2}}_{\rm 2D}$, with the middle row unexcited.
The first probe emphasizes the ground-state region, while the second has a higher mean energy.
Their reconstructed densities of states retain distinct features despite the dense spectrum; see Figs.~\ref{fig:fig5}b and \ref{fig:fig5}c.
These features arise from the probe-dependent eigenstate weights.
To characterize the spatial structure of the filtered states, we evaluate the density wave order.
For the wavevectors considered here, the spatial pattern $s_{\mathbf q,j}=\cos(\mathbf q\cdot\mathbf R_j)$, evaluated on the integer lattice coordinates $\mathbf R_j$, takes values $\pm1$.
The mode at $\mathbf q=(\pi,\pi)$ measures checkerboard order, whereas $\mathbf q=(\pi,0)$ and $(0,\pi)$ probe stripe-like density modulations along each of the two lattice directions.
Changing the probe reverses the sign of the $(\pi,\pi)$ and $(\pi,0)$ density modes, while the $(0,\pi)$ component remains positive.
The experimentally reconstructed modes largely follow the ordered phase of the probe state across the spectrum, with a slight decay of the order toward higher energies; see Fig.~\ref{fig:fig5}d.

\label{sec:conclusion}
In summary, we have demonstrated a hardware-efficient approach to quantum phase estimation on a Rydberg quantum simulator.
Our protocol reconstructs the complex Loschmidt echo to probe many-body spectra in systems with up to 19 atoms.
Extending the reconstruction to generalized Loschmidt echoes additionally provides access to energy-filtered observables.
The sign reversal of the transverse-drive contribution to the energy connects spectral information to coherent many-body structure.
The acquisition time range in the current experiment was chosen subject to the available shot budget and was not limited by dephasing.
Faster repetition rates~\cite{Li2026Fast,Evered2026Highfidelity} and longer coherent evolution increase the spectral resolution and open access to spectral regions with smaller gaps near the phase transition and more precise energy estimation.
Alternatively, different initial probe states with controlled mean energies and sufficient overlap in the target window would enable fine-grained scans of its spectral features.

In principle, our approach provides access to the full state-independent DOS by appropriately sampling from a range of initial product states.
Our results thus open a route to studying many-body systems with a focus on their spectral features that are not or only indirectly accessible by other means.
For example, one may leverage energy-resolved observables to sample states weighted by the Boltzmann factor and obtain filtered observables at a finite temperature~\cite{Lu2021Filtering}.
Furthermore, an interesting extension would be frustrated two-dimensional Rydberg arrays, where quantum fluctuations can easily tip the energy balance between competing phases.
In particular, resolving the energy contributions could provide insight into the kagome Rydberg lattice, where quantum fluctuations are predicted to stabilize a stripe phase between disordered and nematic regimes~\cite{Samajdar2021Quantum}.
Another interesting target would be the Fermi-Hubbard model, where one may consider a plaquette ground state as an initial state as proposed in Ref.~\cite{Cavallar2026PhaseSensitive}.
Finite-energy spectroscopy of this model is motivated, for example, by pseudogap physics~\cite{Kendrick2026Pseudogap, Chalopin2026Observation}, where our approach could provide access to spin and pairing correlations within selected energy windows.
We emphasize that the native analog evolution in our protocol enables long-time evolution without the overhead of compiling the dynamics into product formulas~\cite{Childs2021Theory}, which translates to a narrow filter width and high energy resolution.
This is particularly relevant in regimes where established classical methods, such as matrix-product-state simulations in two dimensions, lose accuracy at long evolution times~\cite{Haghshenas2026Digital, Andersen2026Precision, Leviatan2026Resolving}.
Our results demonstrate statistical quantum phase estimation as a promising avenue to uncover the connection between microscopic and spectral features in quantum many-body systems.

\begin{acknowledgments}
We thank Taylor Briggs for help in setting up the \SI{408}{\nano\meter} and \SI{496}{\nano\meter} laser systems.
We thank Sirui Lu and Yilun Yang for helpful discussions.
We acknowledge funding from the Max Planck Society (MPG), the Deutsche Forschungsgemeinschaft (DFG, German Research Foundation) under Germany's Excellence Strategy--EXC-2111--390814868, and through JST-DFG2024: Japanese-German Joint Call for Proposals on ``Quantum Technologies'' (Japan-JST-DFG-ASPIRE 2024) under DFG Grant No.~554561799, from the Munich Quantum Valley initiative as part of the High-Tech Agenda Plus of the Bavarian State Government, and from the BMFTR through the programs MUNIQC-Atoms and MAQCS.
This publication has also received funding under the Horizon Europe programme HORIZON-CL4-2022-QUANTUM-02-SGA via the project 101113690 (PASQuanS2.1).
J.Z.~acknowledges support from the BMFTR through the program ``Quantum technologies---from basic research to market'' (SNAQC, Grant No.~13N17337).
B.F.S.~and J.I.C.~acknowledge funding from the German Federal Ministry of Research, Technology and Space (BMFTR) through the funded project ALMANAQC, grant number 13N17236 within the research program ``Quantum Systems''.
O.L.~acknowledges support from the Rothschild and CHE Quantum Science and Technology fellowships, as well as funding from the European Union’s Horizon Europe research and innovation programme under the Marie Skłodowska-Curie grant agreement No.~101276098.
H.T.~and M.A.~acknowledge funding from the International Max Planck Research School (IMPRS) for Quantum Science and Technology.
M.A.~acknowledges support through a fellowship from the Hector Fellow Academy.

Competing interests: J.Z.~is a co-founder and shareholder of PlanQC GmbH.

\end{acknowledgments}

\bibliography{bibliography}
\clearpage
\renewcommand{\appendixname}{Supplementary Information}
\appendix

\begin{center}
    {\large\bfseries Supplementary Information for\par}
    \vspace{0.5em}
    {\large\bfseries Quantum phase estimation on a Rydberg quantum simulator\par}
\end{center}
\vspace{1em}

\section{Experimental platform and pulse sequence}
\label{app:experimental_platform}

We trap single $^{88}\mathrm{Sr}$ atoms in optical tweezers generated by a spatial light modulator and crossed acousto-optic deflectors.
The state $\ket{0}$ is encoded in the metastable ${}^3\mathrm{P}_0$ clock state and $\ket{1}$ in the $n=47$, ${}^3\mathrm{S}_1$, $m_J=-1$ Rydberg state~\cite{Madjarov2020,tao2025universal}.
The \SI{813}{\nano\meter} tweezers trap the ground and clock states near magic conditions while strongly shifting the clock-to-Rydberg transition.
Selective release of one sublattice of the 1D chain in AOD tweezers enables site-resolved resonant rotations while the atoms in SLM traps experience a strongly detuned pulse and remain in the clock state.
The local rotations and the analog evolution use the same clock-to-Rydberg laser with $\Omega/(2\pi)=2.3$--$\SI{3.2}{\mega\hertz}$; see Table~\ref{tab:dataset_physical_parameters} for the value used in each dataset.
This prepares the $\ket{Z_2}$ probe digitally.

For the ITE branches, we adjust the pulse area and its phase relative to the analog drive to implement the local $R_y(\beta)$ rotations. 
For the initially excited spins, the rotation angle $\beta$ is related to the ITE step by $\tau\simeq\beta/\Omega$ to leading order, as derived in SI Sec.~\ref{app:ite_derivation}.
Reversing its sign yields the opposite short ITE step.
This operation implements a unitary that approximates normalized ITE to first order in $\tau$ in the stated blockade approximation.

Each shot consists of loading, rearranging, and cooling the atoms near their radial motional ground state to a mean occupation of $\bar{n} \approx 0.1$, followed by preparation in the clock state. 
Following the transfer to the clock state, a fast erasure image using counterpropagating pulses detects residual ground-state population~\cite{Scholl2023,tao2025universal}. 
It typically flags about $1\%$ of the atoms, and we reject the corresponding shots. Assuming an erasure-detection fidelity of $0.99$, this leaves a residual electronic ground-state population of about $10^{-4}$ per atom.
The mobile AOD tweezers then move the selected atoms into the required one- or two-dimensional configuration.
Independent control of the SLM and AOD trapping light allows us to shift the Rydberg resonance of atoms in SLM traps during state preparation and to turn off the trapping light for all atoms during analog evolution.

We prepare the probe with or without the auxiliary rotations, then apply native analog evolution under Eq.~\eqref{eq:rydberg_hamiltonian}.
Rydberg excitations are removed by ionization before fluorescence readout.
We repump the ${}^3\mathrm{P}_2$ manifold into the ground state to identify Rydberg-state decay, and reject shots with detected population; see Ref.~\cite{tao2025universal}.
Finally, slow fluorescence imaging measures the remaining ${}^3\mathrm{P}_0$ population.
We also discard shots with adjacent undetected atoms, which can indicate a blockade violation or atom loss.

\subsection{Physical model}
The one- and two-dimensional geometries have nearest-neighbor spacing \SI{2.6}{\micro\meter}, corresponding to $V_{\rm nn}/(2\pi)=\SI{35}{\mega\hertz}$.
We describe the array by an interaction graph $G=(\mathcal V,\mathcal E)$, whose vertices label the atoms and whose edges connect all interacting pairs, with weights $V_{ij}=C_6/|\mathbf r_i-\mathbf r_j|^6$.
Let $\mathcal N(i)$ denote the nearest neighbors of site $i$ in the array, and let $P_i=1-n_i$ be the projector for that site onto the state $\ket{0}_i$.
For $V_{\rm nn}\gg\Omega,|\Delta|$, the model is approximated by the PXP Hamiltonian with interaction tails,
\begin{equation}
    H =
    \tfrac{\Omega}{2}\sum_{i\in\mathcal V} X_i\prod_{j\in\mathcal N(i)}P_j
    -\Delta\sum_{i\in\mathcal V}n_i +\sum_{\{i,j\}\in\mathcal E}V_{ij}n_i n_j,
    \label{eq:pxptails_hamiltonian}
\end{equation}
effectively restricting the dynamics to the subspace without simultaneous excitations on nearest-neighbor sites.
This expression applies to both one- and two-dimensional arrays.
For an open one-dimensional chain, the constrained flip reduces to $P_{i-1}X_iP_{i+1}$, with $P_0=P_{N+1}=1$.
For each dataset, we calibrate $\Omega$, $\Delta$, and the pulses used to prepare the evolution with and without the ITE.

\begin{figure}[t]
    \centering
    \includegraphics[width=\columnwidth, trim={0mm 0mm 0mm 1.5mm},clip]{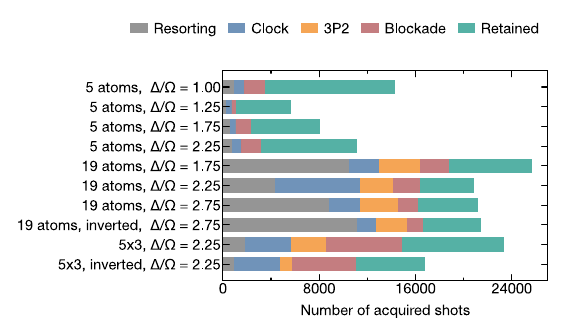}
    \caption{\label{fig:fig_S_postselection} \textbf{Postselection accounting.}
    Number of shots used for the datasets presented. Colored segments show the mutually exclusive numbers rejected by failed resorting, imperfect clock-state preparation, population in the ${}^3\mathrm{P}_2$ manifold, and the blockade criterion; the final segment gives the retained shots. For acquisitions without a separate ${}^3\mathrm{P}_2$ image, this category is absent.}
\end{figure}

\subsection{Postselection of shots}
We apply several different postselection criteria sequentially, with each rejected shot assigned to the first criterion that it fails.
We first reject shots with imperfect resorting. We then reject shots with detected ground-state population in the erasure image, which indicates imperfect clock-state preparation.
For datasets with more than five atoms, a separate ${}^3\mathrm{P}_2$ erasure image additionally identifies population in that manifold, and we reject the corresponding shots.
Finally, we discard configurations with adjacent undetected atoms because they are incompatible with the strong nearest-neighbor blockade constraint or indicate atom loss.
The remaining configurations constitute the retained dataset that we analyze. 
We show the number of rejected shots for each dataset in Fig.~\ref{fig:fig_S_postselection}.

\subsection{Physical parameters}
\label{app:parameters}

Table~\ref{tab:dataset_physical_parameters} summarizes the Hamiltonian parameters and auxiliary imaginary times for each experimental dataset.
We use the Gaussian energy filter
\begin{equation}
    P_\delta(E) =\exp\!\left[-\frac{(H-E)^2}{4\delta^2}\right],
\end{equation}
where $\delta$ is quoted in total-energy angular-frequency units.
For all state-dependent densities of states we use
$\delta_{\mathrm{DOS}}/(2\pi)=0.80\,\si{\mega\hertz}$.
For the energy-filtered observables, the five-atom observables in Fig.~\ref{fig:fig3}b--e, the 19-atom staggered magnetizations, and the $5\times3$ density modes, we use $\delta/(2\pi)=1.13\,\si{\mega\hertz}$.
The energy decomposition in Fig.~\ref{fig:fig3}f uses a broader filter, $\delta/(2\pi)=2.03\,\si{\mega\hertz}$.
The same finite-time Gaussian filter is applied both to the experimental traces and to the corresponding sampled numerical model.

While the rotation angle to implement the ITE can be calculated, see SI Sec.~\ref{app:ite_derivation}, we need to account for the effect of a finite rise time of the acousto-optic modulator in the experiment. 
We therefore determine $\tau$ from the mean Rydberg occupation of the locally targeted atoms after state preparation, using the calibrated Rabi frequency $\Omega$, as
$\tau=\Omega^{-1}\arccos[(\bar n_-+\bar n_+)/\bar n_0-1]$,
where $\bar n_\alpha$ is the mean of $\langle n_i\rangle$ over the targeted sites at $t=0$ in branch $\alpha\in\{0,-,+\}$.

\begin{table*}[ht]
    \centering
    \setlength{\tabcolsep}{4pt}
    \begin{tabular}{llrrrr}
        \toprule
        Geometry
        & Initial state
        & $\Delta/\Omega$
        & $\Omega/(2\pi)$
        & $\Delta/(2\pi)$
        & $\tau/t_{\pi}$ \\
        &
        &
        &
        (\si{\mega\hertz})
        & (\si{\mega\hertz})
        & \\
        \midrule

        1D, $N=5$
        & $\ket{Z_2}$
        & 1.00 & 3.13 & 3.13 & 0.110 \\

        1D, $N=5$
        & $\ket{Z_2}$
        & 1.25 & 3.00 & 3.75 & 0.081 \\

        1D, $N=5$
        & $\ket{Z_2}$
        & 1.75 & 3.15 & 5.51 & 0.163  \\

        1D, $N=5$
        & $\ket{Z_2}$
        & 2.25 & 3.13 & 7.04 & 0.124  \\

        1D, $N=19$
        & $\ket{Z_2}$
        & 1.75 & 2.90 & 5.08 & 0.108  \\

        1D, $N=19$
        & $\ket{Z_2}$
        & 2.25 & 3.00 & 6.75 & 0.102  \\

        1D, $N=19$
        & $\ket{Z_2}$
        & 2.75 & 2.83 & 7.78 & 0.126  \\

        1D, $N=19$
        & $\ket{\overline{Z_2}}$
        & 2.75 & 2.78 & 7.65 & 0.121 \\

        $5\times3$
        & $\ket{Z_2}_{\mathrm{2D}}$
        & 2.25 & 2.27 & 5.11 & 0.097 \\

        $5\times3$
        & $\ket{\overline{Z_2}}_{\mathrm{2D}}$
        & 2.25 & 2.33 & 5.24 & 0.166 \\

        \bottomrule
    \end{tabular}
    \caption{%
        Physical Hamiltonian and ITE parameters for the datasets used
        in the analysis. Frequencies are quoted as ordinary frequencies in
        \si{\mega\hertz}. The Rabi $\pi$-pulse time is defined separately for each
        dataset as
        $t_{\pi} = \pi/\Omega = [2\Omega/(2\pi)]^{-1}$.
        We report the auxiliary imaginary times as fractions of $t_{\pi}$.
    }
    \label{tab:dataset_physical_parameters}
    \vspace{8mm}
\end{table*}

\section{Details on classical postprocessing}
\label{app:postprocessing}

\subsection{Normalization}
After measurement and postselection we obtain bitstrings of computational-basis states as a function of the evolution time for each of the three branches.
Their amplitudes at each $t_k$ refer to the same evolution time.
We take three times as many shots for the prepared initial state without any analog evolution.
For a probe state $\ket{z}$, the postselected return probability in branch $\alpha\in\{0,-,+\}$, corresponding respectively to the main, $-\tau$, and $+\tau$ branches, is
\begin{equation}
    p_z^{(\alpha)}(t)
    =
    \frac{N_z^{(\alpha)}(t)}
         {N_{\mathrm{post}}^{(\alpha)}(t)},
\end{equation}
where $N_z^{(\alpha)}$ is the number of shots returning to the probe state and $N_{\mathrm{post}}^{(\alpha)}$ is the number of retained shots.
We then normalize the return amplitudes with the state-preparation fidelity to mitigate errors in state-preparation and readout that are common to all three branches:
\begin{equation}
    A_\alpha(t_k)
    =
    \sqrt{
        \frac{p_z^{(\alpha)}(t_k)}
             {p_z^{(0)}(0)}
    }.
    \label{eq:probability_to_amplitude}
\end{equation}
As explained in the main text, the local rotations prepare normalized approximations to the two auxiliary ITE states.
They reproduce the change of the normalized state under short ITE, but not the change of its norm.
To recover the corresponding unnormalized magnitudes, we multiply them by the corresponding norms, while keeping corrections for state-preparation and readout errors:
\begin{align}
    R_-(t_k)&=\mathcal N_- A_-(t_k),\\
    R_+(t_k)&=\mathcal N_+ A_+(t_k).
\end{align}
The normalization factors are
\begin{equation}
    \mathcal N_\pm
    =\left\|e^{\pm\tau H}\ket{z}\right\|
    =\sqrt{\bra{z}e^{\pm 2\tau H}\ket{z}}.
\end{equation}
To leading order in $\tau$, they are determined by the mean energy of the probe state,
\begin{equation}
    E_z=\bra{z}H\ket{z},
\end{equation}
which can be calculated efficiently for the probe states considered here.
Expanding the normalization factors gives
\begin{equation}
    \mathcal N_\pm
    =1\pm\tau E_z+\mathcal O(\tau^2)
    =e^{\pm\tau E_z}+\mathcal O(\tau^2).
\end{equation}
Hence, $R_-$ and $R_+$ provide the magnitudes $|g(t-i\tau)|$ and $|g(t+i\tau)|$, respectively.

\subsection{Phase velocity}
The phase velocity is reconstructed as
\begin{equation}
    \dot{\varphi}(t_k)
    =
    \frac{
        \log R_-(t_k)-\log R_+(t_k)
    }{2\tau},
    \label{eq:numerical_phase_velocity}
\end{equation}
following Ref.~\cite{Yang2024}.
The norm factors thus provide a contribution $-E_z$ to the phase velocity at the initial time and thereby establish the energy reference of the reconstruction.
We then numerically integrate the phase velocity to obtain the phase
\begin{equation}
    \varphi(t_k)
    =
    \varphi(0)
    +
    \int_0^{t_k}\di t'\,\dot{\varphi}(t'),
    \qquad
    \varphi(0)=0.
    \label{eq:numerical_phase_integration}
\end{equation}
Together with the main branch, this reconstructs the full Loschmidt echo in Eq.~\eqref{eq:loschmidt_amplitude}.

\subsection{Spectral analysis}
For a classical probe state $\ket{\psi_0}=\ket{z}$, the Loschmidt echo has the spectral decomposition
\begin{equation}
    g(t)=\sum_n
    |\langle E_n|z\rangle|^2e^{-iE_nt}.
\end{equation}
The Gaussian-broadened probe-state-dependent density of states is
\begin{equation}
    \rho_{\psi_0}(E,\delta)
    =
    \sum_n
    |\langle E_n|z\rangle|^2
    \exp\left[
        -\frac{(E-E_n)^2}{2\delta^2}
    \right].
    \label{eq:postprocessing_projected_dos}
\end{equation}
The energies are absolute in the sense that they are referenced to the Hamiltonian convention in Eq.~\eqref{eq:rydberg_hamiltonian}. The density of states is obtained from the Loschmidt echo via a Fourier transformation:
\begin{equation}
    \rho_{\psi_0}(E,\delta)
    =
    \frac{\delta}{\sqrt{2\pi}}
    \int_{-\infty}^{\infty}\di t\,
    e^{-\delta^2t^2/2}
    e^{iEt}g(t).
    \label{eq:postprocessing_fourier}
\end{equation}
Hermiticity implies $g(-t)=g(t)^*$.
Let $\rho_{\psi_0}^{(T)}$ denote the spectrum reconstructed from finite-time data.
For the presentation, each curve is then normalized over the reconstruction grid $\mathcal G$:
\begin{equation}
    \rho_{\psi_0}^{\mathrm{plot}}(E,\delta)
    =
    \frac{
        \rho_{\psi_0}^{(T)}(E,\delta)
    }{
        \displaystyle
        \int_{\mathcal G}\di (E'/N)\,
        \rho_{\psi_0}^{(T)}(E',\delta)
    }.
    \label{eq:plotted_dos_normalization}
\end{equation}
The integral runs over the energy per site $E'/N$ in units of \si{\radian\per\micro\second}, which sets the scale for $\rho_{\psi_0}$.
It is evaluated using the same finite energy grid used in the reconstruction.
The figures always show the normalized probe-state-dependent density of states; we drop the superscript ``plot'' for display purposes.

We suppress late-time data above a time $t_{\mathrm{cut}}$ with a window function,
\begin{equation}
    w_{\mathrm{exp}}(t)=
    \begin{cases}
        1, & |t|\leq t_{\mathrm{cut}},\\[1mm]
        \epsilon^{u(t)^p}, & t_{\mathrm{cut}}<|t|\leq t_{\max},\\[1mm]
        0, & |t|>t_{\max},
    \end{cases}
    \label{eq:soft_exponential_taper}
\end{equation}
where $u(t)=(\lvert t\rvert-t_{\mathrm{cut}})/(t_{\max}-t_{\mathrm{cut}})$, $p=2$, and $\epsilon=0.05$.
Here $t_{\max}$ is the last retained evolution time.
Where the window function is applied, we set $t_{\mathrm{cut}}=\min(\tfrac{1}{3}\,\si{\micro\second},t_{\max})$, independently of the Gaussian width.
The window function reduces ringing of the spectrum due to hard cutoffs of the time signal.
Where applied, this window multiplies the Gaussian factor in the finite-time DOS reconstruction. For the 19-atom staggered magnetization and the two-dimensional density modes, we apply the Gaussian filter over the measured time interval without this additional window.

\subsection{Extracting eigenenergies}
The spectra then allow us to extract eigenenergies of interest.
The reconstruction uses angular-frequency units, consistent with Eq.~\eqref{eq:rydberg_hamiltonian}.
Figures use either the energy $E/(N\Omega)$, the probe-referenced energy $(E-E_z)/(N\Omega)$, or the effective band separation $(E_{\mathcal C_1}^{(z)}-E_{\rm GS})/(N\Omega)$, as indicated on their axes.
The subtraction of $E_z$ is performed only after the absolute spectrum has been reconstructed.
\begin{figure*}[t]
    \centering
    \includegraphics[width=\textwidth, trim={0mm 0mm 0mm 1.5mm},clip]
    {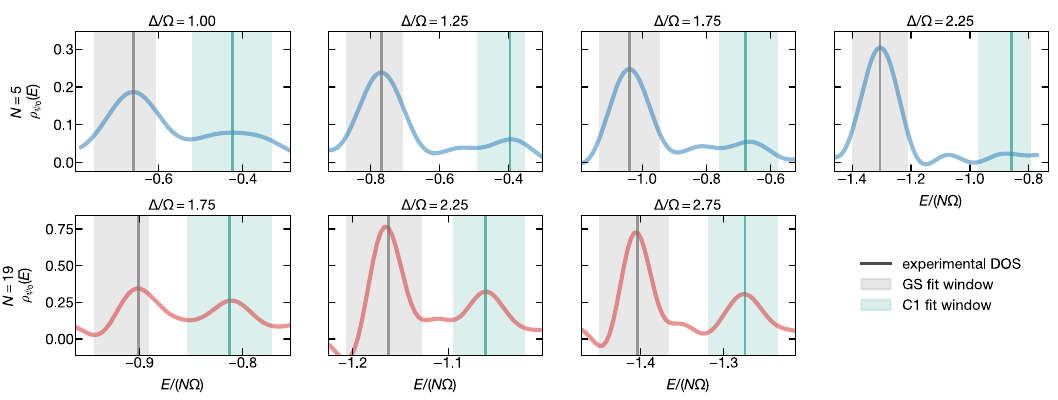}
    \caption{
        \label{fig:GS_C1_fit}
        \textbf{Fitting ground-state and excited-band peaks.}
        Experimental state-dependent densities of states for the $N=5$ chains (upper row, blue) and the $N=19$ chains (lower row, red).
        Each panel is labeled by its detuning.
        The gray and teal background regions indicate the energy-grid points
        included in the ground-state and first-excited-band fits,
        respectively.
        Gray and teal vertical lines show the fitted centers obtained from the
        original experimental reconstruction.
        The corresponding peak-position intervals are obtained by repeating
        the complete reconstruction and fitting procedure for every bootstrap
        replica and are reported in Fig.~\ref{fig:fig2}f,g.
    }
\end{figure*}
We estimate the ground-state energy from the reconstructed experimental
DOS by fitting a Gaussian plus a constant background within the gray
shaded energy window shown in Fig.~\ref{fig:GS_C1_fit}.
The fitted center defines $E_{\mathrm{GS}}$.
For comparison with the observed ground-state energy lowering, we write
$H=H_0+H_X$, where $H_0=H_\Delta+H_V$ is diagonal in the computational basis and
$H_X=\Omega O_X/2$ is the projected transverse drive.
Second-order perturbation theory then gives
\begin{equation}
    E_{\rm GS}-E_{Z_2}
    \simeq
    -\sum_{m\neq Z_2}
    \frac{|\langle m|H_X|Z_2\rangle|^2}
    {E_m^{(0)}-E_{Z_2}^{(0)}}.
    \label{eq:second_order_shift}
\end{equation}
Here, $E_m^{(0)}$ is the eigenenergy of the computational-basis state $\ket{m}$ under $H_0$.
When the longer-ranged interaction tails are neglected, the relevant intermediate states have an energy denominator of $\Delta$.
Since there are $(N+1)/2$ allowed single-spin processes in a long chain, this yields
$E_{\rm GS}-E_{Z_2}\simeq-[(N+1)/2]\,\Omega^2/(4\Delta)$, as quoted in the main text.

We define the first excited band $\mathcal C_1$ as the eigenstates that connect adiabatically to the configurations with one Rydberg excitation fewer than the N\'eel-state probe with $(N+1)/2$ excitations in the limit $\Omega\to0$.
There are $\binom{(N+3)/2}{(N-1)/2}$ configurations in the PXP model with $(N-1)/2$ excitations, so that, ordering the eigenenergies as $E_0<E_1<\cdots$, $\mathcal C_1$ contains six eigenstates for $N=5$ and 55 eigenstates for $N=19$.
This band remains separated from the ground state and the next band for all values of $\Omega$ in the datasets.
To extract the experimental position of this band, we fit a Gaussian plus a constant background within the teal-shaded energy window in Fig.~\ref{fig:GS_C1_fit}.
The fitted center defines $E_{\mathcal C_1}$.
For comparison, we use the numerical mean energy of the band weighted by the probe-state overlap:
\begin{equation}
    E_{\mathcal C_1}^{(z)}
    =
    \frac{
        \displaystyle
        \sum_{n\in\mathcal C_1}
        |\langle E_n|z\rangle|^2E_n
    }{
        \displaystyle
        \sum_{n\in\mathcal C_1}
        |\langle E_n|z\rangle|^2
    }.
    \label{eq:projected_cluster_centroid}
\end{equation}
The separation $E_{\mathcal C_1}^{(z)}-E_{\rm GS}$ provides a numerical reference to compare against the experimentally measured peak separation, but should not be interpreted as the gap to the first excited eigenstate.
The superscript $(z)$ is not included in the main text to simplify notation.

\subsection{Uncertainties}
For each branch and evolution time, we generate bootstrap return counts from a binomial distribution using the measured return probability and the corresponding number of retained shots.
For basis-state-resolved observables, the complete bitstring histogram at each branch and time point is resampled.
Each bootstrap replica is propagated through amplitude reconstruction, phase integration, filtering, normalization, and observable evaluation.
The ground-state and first-band maxima and their local fitting intervals are determined independently for every replica.
We form $E_{\mathcal C_1}-E_{\rm GS}$ within each replica before calculating the interval, thereby retaining the covariance between the two fitted peak positions.
The error bars in Fig.~\ref{fig:fig2}f,g are the central $68\%$ percentile intervals of these distributions.
They quantify finite-shot uncertainty conditional on the postselection, spectral filter, search intervals, and fitting model.

All reconstructed uncertainties use $1000$ paired bootstrap replicas and central $68\%$ intervals.
These intervals describe shot noise conditional on the postselection and do not quantify calibration errors in the Hamiltonian parameters.

In addition, calibration errors can affect the energy offset of the reconstructed spectrum.
This is because the spectrum is referenced to the mean probe-state energy $E_z$, calculated from the calibrated Hamiltonian.
For the N\'eel probe with $(N+1)/2$ excitations, a detuning uncertainty $\sigma_\Delta$ contributes an uncertainty $\sigma_{E_z,\Delta}=[(N+1)/2]\sigma_\Delta$ to this offset, with the interaction parameters held fixed.
We do not include this contribution in the reported statistical uncertainties.

\subsection{Sampled model}
For each dataset, we construct the PXP Hamiltonian with interaction tails using the independently calibrated geometry, $\Omega$, $\Delta$, and $V_{ij}$.
We evaluate the numerical dynamics on the experimental time grid.
The resulting finite-time numerical comparison is referred to as the \textit{sampled model}.

For the probe-state-dependent DOS, we calculate the return-amplitude magnitudes of the main and ITE branches and reconstruct the phase before Fourier filtering.
For the energy-filtered observables, we directly calculate the complex basis-state amplitudes $g_b(t)=\bra{b}e^{-iHt}\ket{\psi_0}$ and apply the same time-domain filter as in the experimental reconstruction.
The model observables are evaluated using the full blockaded basis and normalized by the norm of the filtered state.
The sampled-model observable curves therefore account for the experimental time grid and finite-time filtering, but do not include errors from state preparation or readout.

\section{Wavefunction reconstruction}
\label{app:state_reconstruction}

\begin{figure*}[t]
    \centering
    \includegraphics[width=\textwidth, trim={0mm 0mm 0mm 4mm},clip]
    {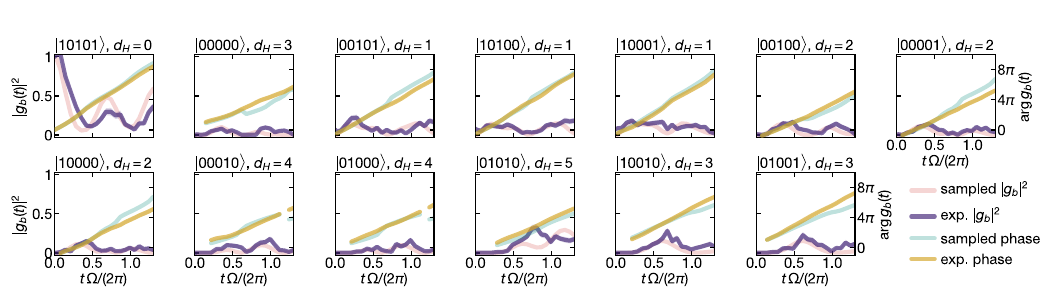}
    \caption{
        \label{fig:basis_state_diagnostics}
        \textbf{State-resolved reconstruction diagnostics for five atoms at
        $\Delta/\Omega=1$.}
        All 13 computational-basis states allowed by the nearest-neighbor
        blockade are shown, labeled by their Hamming distance $d_H$ to the probe.
        Experimental and simulated state probabilities
        $|g_b(t)|^2$ (left axis) and phases $\arg g_b(t)$ (right axis) are shown together for each state.
        The relative phases are fixed by the analytic Hamming-distance anchor including the initial-time correction; see Eq.~\eqref{eq:corrected_generalized_phase}.
        Shaded regions denote pointwise 68\% shot-bootstrap confidence
        intervals.
        Phase values are omitted where the corresponding amplitude is too small for reliable reconstruction.
    }
\end{figure*}

To reconstruct the energy-filtered observables, we analyze the measurement histograms for each computational-basis configuration rather than retaining only the return outcome.
For the five-atom N\'eel probe, the required amplitudes are
\begin{equation}
    g_b(t)=\bra{b}e^{-iHt}\ket{Z_2}.
\end{equation}
The nearest-neighbor-blockaded five-site chain has 13 allowed basis states.
Fig.~\ref{fig:basis_state_diagnostics} compares their measured probabilities and reconstructed complex amplitudes with the sampled model.
Phase reconstruction and its analytic integration constants are described in SI Secs.~\ref{app:ite_derivation} and~\ref{app:phase_anchor}.

In principle, the reconstruction of the complex amplitudes $g_b(t)$ fully specifies the state $\ket{\psi(t)}=\sum_b g_b(t)\ket{b}$.
Hence, one can reconstruct additional properties such as the reduced density matrix $\rho_A(t)=\operatorname{Tr}_{B}[\ket{\psi(t)}\bra{\psi(t)}]$ for a subsystem $A$ with complement $B$.
This also determines the second R\'enyi entropy $S_A^{(2)}(t)=-\ln\operatorname{Tr}[\rho_A(t)^2]$.
Applying the same construction to the energy-filtered state in Eq.~\eqref{eq:filtered_state} would then characterize entanglement within selected spectral windows.
However, we emphasize that these extensions require reliable relative phases from sufficient shots of the populated computational-basis states, which is not generally efficient as the system size grows.

\section{Derivation of the phase reconstruction protocol}
\label{app:ite_derivation}

\subsection{Approximating the ITE with local rotations}
We derive how a short imaginary-time evolution (ITE) step allows us to reconstruct the phase from measured amplitudes, following Refs.~\cite{Yang2024,Schiffer2025}.
ITE shifts the time argument of $g$ into the complex plane, where the Cauchy--Riemann equations relate changes in its magnitude to changes in its phase.
Writing $g(t+i\tau)=r(t,\tau)e^{i\varphi(t,\tau)}$, these equations are
\begin{equation}
    \partial_t\log r=\partial_\tau\varphi,
    \qquad
    \partial_\tau\log r=-\partial_t\varphi,
\end{equation}
locally where $g$ is nonzero and the phase can be continuously defined.
The ITE step thus realizes an effective Wick rotation, and approximating the imaginary-time derivative by finite differences gives the real-time phase velocity:
\begin{equation}
    \frac{\di\varphi(t)}{\di t}
    = \frac{\log|g(t-i\tau)|-\log|g(t+i\tau)|}{2\tau}
    +\mathcal O(\tau^2).
    \label{eq:phase_reconstruction}
\end{equation}
Here $g(t\pm i\tau)$ are the amplitudes for unnormalized ITE.
The normalization factors that are needed to recover their magnitudes from the experimentally prepared normalized states are included in SI Sec.~\ref{app:postprocessing}.

Thus the magnitudes of the two ITE branches determine the phase velocity from which we can calculate the phase by numerical integration.
In our implementation, the Rydberg Hamiltonian can be separated into a classical and a quantum part $H = H_c + H_q$, with $H_q\propto\sum_i X_i$.
A classical initial state is an eigenstate of $H_c$.
Writing $H_c\ket{\psi_0}=\zeta_c\ket{\psi_0}$, a first-order Lie--Trotter decomposition gives
\begin{align}
    e^{-H\tau}\ket{\psi_0}
    &=e^{-H_q\tau}e^{-H_c\tau}\ket{\psi_0}+\mathcal O(\tau^2)\nonumber\\
    &=e^{-\zeta_c\tau}e^{-H_q\tau}\ket{\psi_0}+\mathcal O(\tau^2).
\end{align}
This simplification applies to the many-body Rydberg Hamiltonian considered in this paper.

In the limit of strong interactions in the PXP model and a N\'eel probe state, the spins neighboring an excited spin are effectively frozen.
For the one-dimensional N\'eel probe, only initially excited sites contribute allowed spin flips at first order in the ITE step.
Local rotations therefore approximate the
normalized auxiliary states to first order so that we only need to consider single-particle physics to see that the ITE on these excited spins can be implemented using local $Y$ rotations.
We write the Hamiltonian for a single spin
\begin{align}
    h &= \begin{pmatrix}
        0 & \Omega/2\\
        \Omega/2 & -\Delta
    \end{pmatrix}
    =\frac{\Omega}{2}X - \Delta n \nonumber\\
    &= \Omega_\text{eff} [\sin(\theta) X - 2\cos(\theta) n],
\end{align}
where the effective Rabi frequency is $\Omega_\text{eff}=\sqrt{\Delta^2 + \Omega^2}/2$ and $\tan(\theta)=\Omega/\Delta$, i.e.~$\Omega_\text{eff} \sin(\theta) = \Omega/2$ and $\Omega_\text{eff} \cos(\theta) = \Delta/2$.
We parameterize the single-spin state as
\begin{equation}
    \ket{\varphi(\alpha)}=\cos(\alpha/2)\ket{0}+\sin(\alpha/2)\ket{1},
\end{equation}
and define
\begin{equation}
    \langle h\rangle_\alpha=\bra{\varphi(\alpha)}h\ket{\varphi(\alpha)}.
\end{equation}
The required local $Y$ rotation follows by evaluating
\begin{align}
 &(h-\langle h\rangle_\alpha)\ket{\varphi(\alpha)} \nonumber\\
 =\,& \frac{1}{2} [-\Omega \cos(\alpha) + \Delta\sin(\alpha)]
\begin{pmatrix}
    \sin(\alpha/2)\\
    -\cos(\alpha/2)
\end{pmatrix} \nonumber\\
=\,& \Omega_\text{eff} \sin(\alpha-\theta)
\begin{pmatrix}
    0 & 1\\
    -1 & 0
\end{pmatrix}
\ket{\varphi(\alpha)} \nonumber\\
=\,& i\,a(\alpha,\theta)Y\ket{\varphi(\alpha)},
\end{align}
where
\begin{equation}
    a(\alpha,\theta)=\Omega_\text{eff}\sin(\alpha-\theta).
\end{equation}
Consequently, the normalized short ITE step satisfies
\begin{align}
    &\frac{e^{-h\tau}\ket{\varphi(\alpha)}}
    {\left\|e^{-h\tau}\ket{\varphi(\alpha)}\right\|} \nonumber\\
    =\,&\left[1-\tau\left(h-\langle h\rangle_\alpha\right)\right]
    \ket{\varphi(\alpha)}+\mathcal O(\tau^2)\nonumber\\
    =\,&e^{-i a(\alpha,\theta)\tau Y}
    \ket{\varphi(\alpha)}+\mathcal O(\tau^2).
\end{align}
Thus, to first order in $\tau$, normalized ITE can be implemented using a local $Y$ rotation.
With the convention $R_y(\beta)=e^{-i\beta Y/2}$, the corresponding rotation angle is $\beta=2a(\alpha,\theta)\tau$. 
For an initially excited spin, $\alpha=\pi$ and $a(\pi,\theta)=\Omega_{\mathrm{eff}}\sin\theta=\Omega/2$, so the required rotation angle is $\beta=\Omega\tau$ to leading order. 

For the drive alone, we can obtain the rotation angle at finite $\tau$ by evaluating
\begin{equation}
    e^{-\Omega\tau X/2}\ket{1}
    =\cosh(\Omega\tau/2)\ket{1}
    -\sinh(\Omega\tau/2)\ket{0}.
\end{equation}
Comparing this expression with 
\begin{equation}
    R_y(\beta)\ket{1}=\cos(\beta/2)\ket{1}-\sin(\beta/2)\ket{0},
\end{equation}    
we find that the normalized states coincide when $\tan(\beta/2)=\tanh(\Omega\tau/2)$.
The corresponding rotation angle is therefore $\beta=2\arctan[\tanh(\Omega\tau/2)]$, which reduces to the first-order result above for small $\tau$.

In two dimensions, diagonal neighbors interact with strength $V_{\rm diag}=V_{\rm nn}/8$.
Two or four occupied diagonal neighbors then contribute approximately $3.8\Omega$ or $7.6\Omega$ to the interaction shift before including the detuning.
These finite shifts therefore are not a perfect blockade as they do not fully eliminate the first-order ITE amplitudes for initially empty sites without occupied nearest neighbors.
Restricting the rotations for the ITE branches to initially excited sites therefore introduces an additional approximation.
We indirectly quantify the resulting return-phase error below.

\subsection{Accuracy of the two-dimensional auxiliary preparation}
\label{app:2d_preparation_error}

In our two-dimensional arrays presented in Fig.~\ref{fig:fig5} in the main text, diagonal neighbors lie within the conventional blockade radius, motivating auxiliary rotations restricted to initially excited sites.
Their finite interaction strength, $V_{\mathrm{diag}}=V_{\mathrm{nn}}/8$, nevertheless makes this preparation approximate.
We assess its accuracy by exact diagonalization of the $5\times3$ Hamiltonian in the nearest-neighbor-blockaded subspace, retaining all longer-range
interactions and using the parameters in Table~\ref{tab:dataset_physical_parameters}.
We calculate Loschmidt echoes for two preparations: local rotations on occupied sites only, and exact normalized ITE at the same step $\tau$.
The benchmark uses rotation angles $\beta=2\arctan[\tanh(\Omega\tau/2)]$.

We let $v_{\mathrm{rot}}(t)$ and $v_{\mathrm{ITE},\tau}(t)$ denote the corresponding central-difference estimates of the phase velocity, and let
\begin{equation}
    v(t)=\operatorname{Im} \left[\frac{\dot g(t)}{g(t)}\right]
\end{equation}
be the exact phase velocity obtained directly from the Loschmidt echo $g(t)=\bra{z}e^{-iHt}\ket{z}$.
For both reconstructions, we fix the initial velocity to its known value $-E_z$, where $E_z=\bra{z}H\ket{z}$, by subtracting the difference
between the reconstructed initial velocity and $-E_z$.
This removes a constant initial offset and leaves the occupied-site-only result unchanged because its initial auxiliary
return magnitudes are equal.
In the following, both reconstructed velocities use this convention.

We distinguish the change caused by replacing exact ITE with local rotations from the error that remains even with exact ITE at finite $\tau$:
\begin{align}
    \delta v_{\mathrm{prep}}(t)
    &=v_{\mathrm{rot}}(t)-v_{\mathrm{ITE},\tau}(t),\\
    \delta v_{\mathrm{step}}(t)
    &=v_{\mathrm{ITE},\tau}(t)-v(t).
\end{align}
Their sum is the total velocity error,
$\delta v_{\mathrm{total}}(t)=\delta v_{\mathrm{prep}}(t)+\delta v_{\mathrm{step}}(t)=v_{\mathrm{rot}}(t)-v(t)$.
The quantities $\delta v_{\alpha}(t)$ are signed, time-dependent errors.
For each $\alpha\in\{\mathrm{prep},\mathrm{step},\mathrm{total}\}$, we define the corresponding dimensionless root-mean-square (RMS) error over the interval $[0,T]$ as
\begin{equation}
    \epsilon_{\alpha}
    =\frac{1}{\Omega}
    \Big[\frac{1}{T}\int_0^T|\delta v_{\alpha}(t)|^2\,\di t\Big]^{1/2}.
    \label{eq:2d_velocity_error}
\end{equation}
Table~\ref{tab:2d_velocity_errors} reports these RMS errors over the acquisition intervals, with $T\approx\SI{0.573}{\micro\second}$ for $\ket{Z_2}_{\mathrm{2D}}$ and $T\approx\SI{0.558}{\micro\second}$ for $\ket{\overline{Z_2}}_{\mathrm{2D}}$.

\begin{table}[t]
    \centering
    \begin{tabular}{lccc}
        \toprule
        Probe
        & $\epsilon_{\mathrm{prep}}$
        & $\epsilon_{\mathrm{step}}$
        & $\epsilon_{\mathrm{total}}$ \\
        \midrule
        $\ket{Z_2}_{\mathrm{2D}}$
        & 0.207 & 0.104 & 0.108 \\
        $\ket{\overline{Z_2}}_{\mathrm{2D}}$
        & 0.692 & 0.373 & 0.428 \\
        \bottomrule
    \end{tabular}
    \caption{%
        Dimensionless root-mean-square (RMS) phase-velocity errors $\epsilon_{\alpha}$, as defined in Eq.~\eqref{eq:2d_velocity_error}, for the two-dimensional probe states.
        The second column (preparation) compares local rotations with exact ITE and the third column (finite-step) compares exact-ITE reconstruction with the exact phase velocity.
        The total column compares the phase velocity obtained using rotations on occupied sites only with the exact phase velocity.
        While signed errors add pointwise, their RMS values do not, because the preparation and finite-step errors partly cancel.
    }
    \label{tab:2d_velocity_errors}
\end{table}

\begin{figure}[b]
    \centering
    \includegraphics[width=\columnwidth]
    {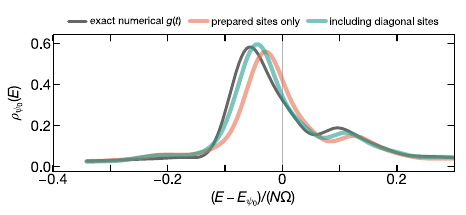}
    \caption{%
        \label{fig:2D_ITE_error}
        \textbf{Effect of diagonal-site rotations on the reconstructed two-dimensional spectrum.}
        Numerical state-dependent DOS for the inverted checkerboard-like state in the $5\times3$ array at $\Delta/\Omega=2.25$.
        The black curve is obtained from the exact numerical Loschmidt amplitude $g(t)$.
        The coral curve uses the experimental ITE protocol, in which the auxiliary rotations are applied only to the initially excited sites.
        The teal curve shows the protocol (not employed in the experiment) that additionally rotates the relevant diagonally coupled sites.
    }
\end{figure}

The contribution of the preparation step is approximately twice the finite-step contribution for both probes.
Their signed errors partly cancel, so their RMS values do not simply add.
The preparation approximation, which includes treating next-nearest neighbors as perfectly blockaded, is therefore of the same order of magnitude as the finite-step reconstruction error.
This benchmark quantifies errors in the phase velocity only.
We compare DOS reconstructions using rotations on occupied sites only with reconstructions that additionally rotate the relevant initially empty sites in the middle row; see Fig.~\ref{fig:2D_ITE_error}.
Having access to full local addressing with an additional beam would allow us to implement this protocol and reduce the error of this additional ITE approximation.

\section{Phase velocity reconstruction for generalized Loschmidt echoes}
\label{app:phase_anchor}

The standard Loschmidt echo $\bra{z}e^{-iHt}\ket{z}$ equals one at time zero, which fixes its initial phase.
For energy-filtered observables we also consider a generalized Loschmidt echo 
\begin{equation}
    g_b(t)=\bra{b}e^{-iHt}\ket{z}=r_b(t)e^{i\varphi_b(t)},
\end{equation}
where $\ket{z}$ is the N\'eel probe state and $\ket{b}$ is a computational-basis state.
The magnitude of the generalized Loschmidt echo $r_b(t)$ starts at zero for $b\neq z$.
We therefore need to fix its initial phase and revisit the accuracy of phase reconstruction at a finite imaginary-time step.

\subsection{Setting the relative phase between basis states}

The same experimental histograms provide amplitude magnitudes and phase velocities for each sufficiently sampled computational-basis configuration $\ket{b}$.
To combine these amplitudes into a filtered state, their phases must be placed in a common convention.
Integration of the measured phase velocity gives the time-dependent phase only up to an integration constant $\varphi_b(0^+)$ for each $b$.
For $b=z$, the known return amplitude $g_z(0)=1$ fixes $\varphi_z(0)=0$.
For $b\neq z$, however, $g_b(0)=0$, so its phase at exactly $t=0$ is undefined.
The required constant is instead the phase of the first nonvanishing term of $g_b(t)$ as $t$ approaches zero from above.

Expanding the propagator of the time evolution yields
\begin{equation}
    g_b(t) = \sum_{n=0}^{\infty}
    \frac{(-it)^n}{n!}
    \langle b|H^n|z\rangle.
\end{equation}
Then, defining the first nonvanishing power of the Hamiltonian as
\begin{equation}
    d_b=
    \min\left\{
        n:\langle b|H^n|z\rangle\neq0
    \right\},
\end{equation}
the leading short-time contribution is
\begin{equation}
    g_b(t)
    =
    \frac{(-it)^{d_b}}{d_b!}
    \langle b|H^{d_b}|z\rangle
    +\mathcal O(t^{d_b+1}).
\end{equation}
The right-hand limit of the phase is then
\begin{equation}
    \varphi_b(0^+)
    =
    \arg\!\left[\langle b|H^{d_b}|z\rangle\right]
    -\frac{\pi d_b}{2}
    \pmod{2\pi}.
    \label{eq:general_phase_anchor}
\end{equation}
These statements apply to arbitrary Hamiltonians provided that $\langle b|H^n|z\rangle\neq0$ for at least one nonnegative integer $n$.
If no such $n$ exists, then $g_b(t)$ vanishes identically and its phase need not be reconstructed.

Interestingly, for the Rydberg Hamiltonian in the PXP limit, see Eq.~\eqref{eq:pxptails_hamiltonian}, the off-diagonal terms are real single-spin flips and the drive can only induce spin flips that remain within the blockaded subspace.
The detuning and interaction terms are diagonal in the computational basis and cannot reduce the number of spin flips required to connect $\ket{z}$ and $\ket{b}$.
Therefore, for two allowed blockade configurations, the minimum number of allowed flips is given by their Hamming distance,
\begin{equation}
    d_b=d_H(b,z).
\end{equation}
At the first nonzero order, only the shortest allowed paths contribute.
For a positive drive, we have
\begin{equation}
    \langle b|H^{d_b}|z\rangle
    =
    N_{\mathrm{sp}}(z,b)
    \left(\frac{\Omega}{2}\right)^{d_b}>0,
\end{equation}
where $N_{\mathrm{sp}}(z,b)$ is the number of shortest allowed paths.
Eq.~\eqref{eq:general_phase_anchor} thus reduces to the simple rule
\begin{equation}
    \varphi_b(0^+)
    =
    -\frac{\pi}{2}d_H(b,z)
    \pmod{2\pi}.
    \label{eq:hamming_phase_anchor}
\end{equation}
The relative phase that we require then only depends on the Hamming distance modulo four:
\begin{equation}
    \begin{array}{c|cccc}
    d_H\bmod 4 & 0 & 1 & 2 & 3\\
    \hline
    \varphi_b(0^+) & 0 & -\pi/2 & \pi & +\pi/2.
    \end{array}
\end{equation}
On a zero-free interval connected to $t=0^+$, the phase of $g_b(t)=\langle b|e^{-iHt}|\psi_0\rangle$ is obtained by integrating the reconstructed phase velocity from the analytic anchor:
\begin{equation}
    \varphi_b(t)
    =
    \varphi_b(0^+)
    +\int_{0^+}^{t}\di t'\,\dot{\varphi}_b(t').
\end{equation}
Continuation across later zeros generally requires a separate prescription to account for possible phase jumps.
The simple analytical rule applies to computational-basis probe states in the hard-blockade limit of the Rydberg model.
The phase anchor depends only on the measured bitstrings and the single-spin-flip structure of the Hamiltonian, making our approach scalable and independent of numerical methods to simulate the time evolution for short times.
This allows us to obtain the generalized Loschmidt echoes $g_b(t)$ used to compute energy-filtered observables without numerically simulating the short-time dynamics. 

\subsection{Initial-time correction of the phase velocity}
\label{app:initial_time_correction}

Here, we calculate a correction term that removes the leading bias arising at early times in the reconstructed phase of generalized Loschmidt echoes.
Within the blockade approximation, this correction brings the error in the phase to $\mathcal{O}(\tau^2)$ for the geometries studied in the experiment.
This systematic error can therefore be reduced by decreasing $\tau$, as is the case for the standard Loschmidt echo.

First, we recall that the Cauchy--Riemann equations provide an estimate of the phase velocity from exact imaginary-time evolution,
\begin{equation}
    \dot{\varphi}_b(t)
    = \frac{\log|g_b(t-i\tau)|-\log|g_b(t+i\tau)|}{2\tau} +\mathcal O(\tau^2),
\end{equation}
provided the magnitude of the Loschmidt echo is nonzero.
Replacing imaginary-time evolution by local rotations introduces an absolute amplitude error of $\mathcal O(\tau^2)$ from the splitting into classical and quantum terms.
To see why this need not give the same order of phase error, we write the logarithmic difference in the previous formula as
\begin{equation}
    \frac{1}{2\tau}\log\frac{|g_b(t-i\tau)|}{|g_b(t+i\tau)|}.
\end{equation}
For the standard Loschmidt echo, where the basis state $b$ equals the initial state $z$, both moduli remain close to one near the initial time.
For a different final configuration in the generalized Loschmidt echo, i.e.~$b\neq z$, both moduli can instead be small, so an error in either modulus can significantly affect their ratio.
For example, if the final state is reached by one spin flip, both moduli are of order $\tau$ when $t\sim\tau$.
Their ratio remains $\mathcal O(1)$, but an absolute amplitude error of $\mathcal O(\tau^2)$ is now a relative error of $\mathcal O(\tau)$.
The logarithm can therefore change by $\mathcal O(\tau)$ and dividing by $2\tau$ gives an $\mathcal O(1)$ error in the phase velocity.
This error acts over an initial time interval of order $\tau$ and can leave an $\mathcal O(\tau)$ error in the integrated phase.
At fixed positive time away from zeros, the two opposite rotations still give a velocity error of $\mathcal O(\tau^2)$.

The rotations used to approximate the ITE reproduce every initially allowed flip to first order but miss energy-dependent corrections at second order.
We consider ideal instantaneous rotations on excited spins with coefficients $\cos(\Omega\tau/2)$ and $\pm\sin(\Omega\tau/2)$.
They reproduce the required spin flips when no additional excitations are allowed in the probe state.
This is the case for the probes considered here within the hard-blockade approximation, including diagonal blockade in two dimensions.
At second order, however, ITE also depends on the classical energy differences between configurations, which these rotations do not capture.
For transition amplitudes that vanish at the initial time, this mismatch can produce the $\mathcal O(\tau)$ integrated phase bias described above.

A single-atom example illustrates how this bias arises concretely.
Let $\ket{z}$ be an initially excited state and $\ket{b}$ the de-excited state, whose classical energy relative to $\ket{z}$ is $e$.
Taking the energy of $\ket{z}$ to be zero, the Hamiltonian in the basis $(\ket{z},\ket{b})$ is
\begin{equation}
    H=\begin{pmatrix}
        0 & \Omega/2\\
        \Omega/2 & e
    \end{pmatrix}.
\end{equation}
The exact phase velocity of $g_b(t)$ is $-e/2$ near the initial time.
For times $t$ of order $\tau$, the rotations above instead give an estimator $\dot{\varphi}_b^{\mathrm{est}}(t)$ with error
\begin{equation}
    \dot\varphi_b^{\mathrm{est}}(t)-\dot{\varphi}_b(t)
    =\frac{e\tau^2}{2(t^2+\tau^2)}+\mathcal O(\tau^2).
\end{equation}
At $t=\tau$, this error approaches $e/4$ as $\tau\to0$, rather than zero.
The contribution decreases as $t/\tau$ grows, but its integral leaves a phase bias:
\begin{equation}
    \int_0^t \frac{e\tau^2}{2(t'^2+\tau^2)} \di t'
    =\frac{e\tau}{2}\arctan\!\left(\frac{t}{\tau}\right).
\end{equation}
For $t\gg\tau$, this bias approaches $\pi e\tau/4$.
The example thus shows how a finite initial velocity error leaves an $\mathcal O(\tau)$ phase bias.

The described bias in the integrated phase can be removed by applying a constant phase offset during postprocessing.
We verified numerically that its coefficient has two contributions for the geometries that we consider,
\begin{equation}
    \kappa_b=\kappa_b^{(1)}+\kappa_b^{(2)}.
    \label{eq:constant_phase_coefficient}
\end{equation}
The first contribution comes from the change in classical energy upon de-exciting an atom.
Let $H_c$ denote the diagonal classical Hamiltonian and $E_b=\bra{b}H_c\ket{b}$ the classical energy of a computational-basis configuration $\ket{b}$.
Removing an excitation from an initially occupied site $i$ changes the energy by $e_i=E_{z\setminus\{i\}}-E_z$.
Let $p_i$ be the ratio of the number of shortest allowed flip sequences from $\ket{z}$ to $\ket{b}$ whose first flip removes the excitation at site $i$ to the total number of shortest allowed sequences between these states.
The first contribution is then
\begin{equation}
    \kappa_b^{(1)}=\frac{\pi d_b}{4}\sum_i p_i e_i,
\end{equation}
where the sum runs over initially occupied sites.

The second contribution is needed when removing one excitation immediately permits an excitation to be added elsewhere.
Let $P_{ij}$ be the fraction of the shortest allowed flip sequences starting with removal of $i$, followed by an excitation at site $j$ that becomes allowed only after removing the excitation at site $i$.
For each complete operation, we average the classical energies of its $d_b+1$ configurations.
Averaging these values over all orderings gives $\overline E_b$.
Another average $\overline E_{b|ij}$ uses only operations beginning with these two flips and only the $d_b-1$ configurations from immediately after them through the final state $b$, i.e.~it excludes $z$ and the first-removal state.
This gives the second contribution
\begin{equation}
    \kappa_b^{(2)}=\frac{\pi d_b(d_b-1)}{4}
    \sum_{i,j}P_{ij}(\overline E_{b|ij}-\overline E_b),
\end{equation}
where the sum includes only removal-addition pairs with $P_{ij}>0$.
If no such pair occurs, then we have $\kappa_b^{(2)}=0$.
This contribution matters, for example, at the endpoints of the $\ket{\overline{Z_2}}$ probe.
For the return amplitude, $\bra{z}e^{-iHt}\ket{z}$, we set $\kappa_z=0$.
Both contributions can be calculated from the bitstrings, classical energies, and counting the shortest paths, without simulating real-time quantum dynamics.

In numerical postprocessing, the prescription is then to multiply each reconstructed generalized Loschmidt echo $g_b(t)$ by $e^{-i\tau\kappa_b}$.
We let $\varphi_b^{\mathrm{corr}}(t)$ be the phase reconstructed from the analytically obtained relative phase $-\pi d_b/2$ between the basis states, including the initial-time correction.
Writing $\dot\varphi_b^{\mathrm{est}}(t)$ for the rotation-based phase velocity estimator with the normalization already included, we obtain
\begin{equation}
    \varphi_b^{\mathrm{corr}}(t)
    =-\frac{\pi d_b}{2}-\tau\kappa_b+\int_{0^+}^{t}\dot\varphi_b^{\mathrm{est}}(t')\,\di t'.
    \label{eq:corrected_generalized_phase}
\end{equation}
For the stated blockade geometries and a fixed reconstruction interval without later amplitude zeros, this offset cancels the integrated $\mathcal O(\tau)$ bias and leaves a phase error of $\mathcal O(\tau^2)$ at fixed positive time.
We note that the preparation error due to finite diagonal interactions in two dimensions is separate from this bias correction; see SI Sec.~\ref{app:2d_preparation_error}.

\section{Reconstruction of energy-filtered observables}
\label{app:filtered_observables}
Applying the Gaussian filter to the initial probe state and normalizing the result defines the energy-filtered state
\begin{equation}
    \ket{\psi_{E,\delta}}
    =
    \frac{1}{\sqrt{\rho_{\psi_0}(E,\delta)}}
    P_\delta(E)\ket{\psi_0}.
    \label{eq:filtered_state}
\end{equation}
The normalization follows from
$\rho_{\psi_0}(E,\delta)
=\bra{\psi_0}P_\delta(E)^2\ket{\psi_0}$.
The computational-basis histograms give the time-dependent populations of the basis states.
Combining them with ITE phase reconstruction and the corrected analytic phase anchor, Eq.~\eqref{eq:corrected_generalized_phase}, gives
\begin{equation}
    g_b(t)=\bra{b}e^{-iHt}\ket{\psi_0}.
\end{equation}
For the real Hamiltonian and real product probes used here, $g_b(-t)=g_b(t)^*$, which provides the negative-time values in the Fourier expression below.
For the Gaussian operator 
\begin{equation}
    P_\delta(E)=\exp\left[-\frac{(H-E)^2}{4\delta^2}\right],    
\end{equation}
the corresponding unnormalized filtered coefficient is
\begin{align}
    \widetilde c_b(E)
    &=\langle b|P_\delta(E)|\psi_0\rangle\nonumber\\
    &=\frac{\delta}{\sqrt{\pi}}
    \int_{-\infty}^{\infty}\di t\,
    e^{-\delta^2t^2}
    e^{iEt}g_b(t).
    \label{eq:filtered_basis_coefficient}
\end{align}
Completeness of the computational basis gives
\begin{equation}
    \sum_b|\widetilde c_b(E)|^2
    =\langle\psi_0|P_\delta(E)^2|\psi_0\rangle
    =\rho_{\psi_0}(E,\delta).
\end{equation}
The normalized coefficients in Eq.~\eqref{eq:filtered_state} are therefore
\begin{equation}
    c_b(E)=\frac{\widetilde c_b(E)}
    {\sqrt{\rho_{\psi_0}(E,\delta)}}.
    \label{eq:normalized_filtered_coefficient}
\end{equation}
Consequently, the normalized filtered state can be written in the computational basis as
\begin{align}
    \ket{\psi_{E,\delta}}
    &=
    \sum_b c_b(E)\ket{b} \nonumber\\
    &=
    \frac{1}{\sqrt{\rho_{\psi_0}(E,\delta)}}
    \sum_b
    \ket{b}\bra{b}P_\delta(E)\ket{\psi_0}.
    \label{eq:filtered_state_basis}
\end{align}
For an operator $O$ with computational-basis matrix elements $O_{bb'}=\langle b|O|b'\rangle$, the energy-filtered expectation value is
\begin{align}
    O_{\psi_0,\delta}(E)
    &=
    \bra{\psi_{E,\delta}}O\ket{\psi_{E,\delta}} \nonumber\\
    &=
    \sum_{b,b'}
    c_b(E)^*O_{bb'}c_{b'}(E).
    \label{eq:filtered_observable_basis}
\end{align}
Diagonal observables therefore depend only on the filtered populations $|c_b(E)|^2$, whereas off-diagonal observables also require the reconstructed relative phases between basis states.
For finite experimental data, we normalize the reconstructed filtered coefficients by the square root of their summed squared magnitude over the retained configurations.
\subsection{Energy-resolved one-point observables and density modes}
For the one-dimensional Rydberg arrays, we evaluate the mean Rydberg density and the projected transverse-drive coherence according to
\begin{align}
    \overline{\langle n\rangle}_E
    &=
    \frac{1}{N}\sum_i\langle n_i\rangle_E,
    \\
    \overline{\langle O_X\rangle}_E
    &=
    \frac{1}{N}
    \sum_i
    \langle P_{i-1}X_iP_{i+1}\rangle_E.
    \label{eq:filtered_one_point_observables}
\end{align}
The first quantity is diagonal in the measurement basis, whereas the second depends on coherences between computational-basis states.

For the one-dimensional chains, with sites numbered $1,\ldots,N$, we define
\begin{equation}
    M_\pi=\frac{1}{N}\sum_{i=1}^{N}(-1)^{i-1}(2n_i-1).
    \label{eq:staggered_magnetization}
\end{equation}
The probe states $\ket{Z_2}=\ket{1010\ldots1}$ and $\ket{\overline{Z_2}}=\ket{0101\ldots0}$ have $M_\pi=+1$ and $-1$, respectively.
The plotted filtered expectation is $\langle M_\pi\rangle_E$.

For the two-dimensional density modes, we use integer lattice
coordinates $\mathbf R_j=(x_j,y_j)$, with $x_j=0,\ldots,4$
along the five-site direction and $y_j=0,\ldots,2$ along
the three-site direction.
The physical positions satisfy
$\mathbf r_j=\mathbf r_0+a\mathbf R_j$,
where $a$ is the lattice spacing.
We define the spatial pattern
$s_{\mathbf q,j}=\cos(\mathbf q\cdot\mathbf R_j)$,
with dimensionless wavevectors $\mathbf q$.
To remove its finite-array mean and normalize the resulting
mode, we use
\begin{equation}
    \bar s_{\mathbf q}
    =
    \frac{1}{N}\sum_j s_{\mathbf q,j},
    \qquad
    f_{\mathbf q,j}
    =
    \frac{s_{\mathbf q,j}-\bar s_{\mathbf q}}
    {\sqrt{N\left(1-\bar s_{\mathbf q}^{2}\right)}}.
    \label{eq:centered_mode_weights}
\end{equation}
For the patterns considered here, for which
$s_{\mathbf q,j}=\pm1$, this convention gives
$\sum_j f_{\mathbf q,j}=0$ and
$\sum_j f_{\mathbf q,j}^2=1$.

The corresponding energy-resolved centered density mode is
\begin{equation}
    m_{\mathbf q}^{\mathrm{cent}}(E)
    =
    \sum_j
    f_{\mathbf q,j}\langle n_j\rangle_E.
    \label{eq:centered_density_mode}
\end{equation}
The wavevector $\mathbf q=(\pi,\pi)$ selects checkerboard order, whereas
$\mathbf q=(\pi,0)$ and $(0,\pi)$ select stripe-like modulations along the two lattice directions.

For the 19-atom magnetization, we retain energies where the experimental filtered norm exceeds $10^{-3}$ of its maximum and the observed configurations carry at least half of the sampled-model filtered weight.
The displayed experimental and model curves also satisfy a relative norm threshold of $10^{-3}$.
For the two-dimensional density modes, we use explicit energy windows.

\subsection{Energy-resolved two-point correlations}

\begin{figure}[t]
    \centering
    \includegraphics[width=\columnwidth, trim={0mm 0mm 0mm 2.5mm},clip]
    {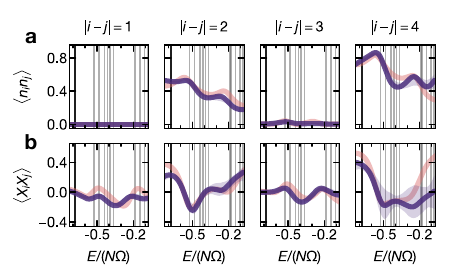}
    \caption{\label{fig:filtered_two_point}
    \textbf{Energy-resolved two-point correlations.}
    Correlators in the five-atom chain at $\Delta/\Omega=1$, averaged over pairs with separation $|i-j|=1,\ldots,4$ (columns).
    \textbf{(a)} Density-density correlations $\langle n_i n_j\rangle_E$, which depend on the filtered basis-state populations.
    The nearest-neighbor correlations vanish because the analysis postselects configurations with no adjacent Rydberg excitations.
    \textbf{(b)} Transverse correlations $\langle X_iX_j\rangle_E$, which probe coherences between basis states differing by flips of spins $i$ and $j$.
    Dark purple curves show the experiment and pink curves the sampled numerical model; shaded regions denote pointwise $68\%$ experimental shot-bootstrap intervals.
    Vertical lines indicate the exact ground-state and excited-state energies.}
\end{figure}

The interaction contribution $H_V$ requires density correlations beyond one-point observables.
Fig.~\ref{fig:filtered_two_point} shows energy-resolved two-point correlations for the five-atom dataset at $\Delta/\Omega=1$.
The density-density correlator is
\begin{equation}
    \langle n_i n_j\rangle_E
    =
    \sum_b |c_b(E)|^2 n_i(b)n_j(b).
\end{equation}
By contrast, the transverse correlator is determined by off-diagonal coherences,
\begin{equation}
    \langle X_iX_j\rangle_E
    =
    \sum_b c_b^*(E)c_{b\oplus i\oplus j}(E),
\end{equation}
where $b\oplus i\oplus j$ denotes the bitstring obtained by flipping sites $i$ and $j$.
Coefficients of configurations outside the allowed subspace are defined to be zero.
Consequently, determining $\langle X_iX_j\rangle_E$ requires relative phases between basis-state amplitudes.
The ITE protocol provides this phase information and allows us to reconstruct coherences within the energy-filtered state.

For the open five-atom chain, we average all pairs at fixed separation,
\begin{equation}
    \overline{\langle O_iO_{i+r}\rangle}_E
    =
    \frac{1}{N-r}
    \sum_{i=1}^{N-r}
    \langle O_iO_{i+r}\rangle_E.
\end{equation}
We use this average with $O_i=n_i$ and $O_i=X_i$ for the density-density and transverse correlations shown in Fig.~\ref{fig:filtered_two_point}, respectively.

\end{document}